\documentclass[aps,pre,onecolumn,floatfix,nofootinbib]{revtex4}

\usepackage{amssymb,amsfonts,amsmath}
\usepackage{graphicx}
\usepackage{dcolumn}
\usepackage{bm}
\usepackage{mathrsfs}
\usepackage{mhchem}
\usepackage{subfigure}
\usepackage{color}

\usepackage{upgreek}

\usepackage{booktabs}
\usepackage{epstopdf}
\usepackage{multirow}

\usepackage{chemarrow}
\usepackage{extarrows}
\usepackage{mathtools}

\usepackage{epsf}
\usepackage{epstopdf}
\newcommand{\be}{\begin{equation}}
\newcommand{\ee}{\end{equation}}
\newcommand{\bea}{\begin{eqnarray}}
\newcommand{\eea}{\end{eqnarray}}

\usepackage[T3,T1]{fontenc}
\DeclareSymbolFont{tipa}{T3}{cmr}{m}{n}
\DeclareMathAccent{\invbreve}{\mathalpha}{tipa}{16}

\newcommand{\dd}{{\mathrm d}}
\newcommand{\ddo}{{\mathrm d}o}
\newcommand{\ddv}{{\mathrm d}v}

\begin{document}

\title{The non-equilibrium thermodynamics of active suspensions}

\author{Pierre Gaspard}
\thanks{ORCID: {\tt 0000-0003-3804-2110}}
\affiliation{ Center for Nonlinear Phenomena and Complex Systems, Universit{\'e} Libre de Bruxelles (U.L.B.), Code Postal 231, Campus Plaine, B-1050 Brussels, Belgium}


\begin{abstract}
Active suspensions composed of self-propelled colloidal particles are considered.  The propulsion of these micrometric particles is generated by chemical reactions occurring by heterogeneous catalysis at their surface and by diffusiophoresis coupling the concentration gradients of the reacting molecular species to the fluid velocity.  By this mechanism, chemical free energy is transduced into mechanical motion.  The non-equilibrium thermodynamics of such active suspensions is developed by explicitly taking into account the internal degrees of freedom of the active particles, which are the Eulerian angles specifying their orientation.  Accordingly, the distribution function of the colloidal particles is defined in the six-dimensional configuration space of their position and their orientation, which fully characterises the polar, nematic, and higher orientational orders in the active system.  The local Gibbs and Euler thermodynamic relations are expressed in terms of the colloidal distribution function, the dynamics of which is ruled by a six-dimensional local conservation equation.  All the processes contributing to the entropy production rate are derived from the local conservation and kinetic equations for the colloids, the molecular species, mass, linear momentum, and energy, identifying their thermodynamic forces -- also called affinities -- and their dissipative current densities.  The non-equilibrium constitutive relations between them are obtained using the Curie symmetry principle and the Onsager-Casimir reciprocal relations based on microreversibility.  In this way, all the coefficients of mechanochemical coupling are completely determined for isothermal, incompressible, dilute suspensions composed of spherical Janus particles on the basis of the interfacial properties between the fluid solution and the solid particles and chemohydrodynamics. The complete expression of the entropy production rate is thus established for such active systems.
\vskip 0.1 cm
{\bf Keywords:} Active matter, chemical reactions, colloidal Janus particles, local conservation laws, entropy production, thermodynamic force, affinity, dissipative current density, mechanochemical coupling, energy transduction.
\end{abstract}


\maketitle

\section{Introduction}
\label{sec:intro}

In active matter, free energy is locally consumed to power the movements of agents, which are spatially distributed in the system.  There exist many different sorts of active matter, including biological tissues, active gels, and ensembles of micro-organisms or synthetic colloids moving in fluids \cite{R10,MJRLPRA13,K13,BDLRVV16}.  These non-equilibrium systems are textured or structured at the intermediate mesoscale between the size of atoms and molecules composing matter and the macroscales of self-organisation into patterns or waves.  These mesotextures or mesostructures have spatial scales typically ranging from $0.1~\mu$m to $100~\mu$m.  In this regard, the agents composing active matter have many internal degrees of freedom, which have crucial functions in the mechanisms generating the transduction of free energy into motility \cite{H77}.  

For instance, micro-organisms obtain nutrients from their environment and their metabolic reaction network supplies their biomolecular motors with chemical or electrochemical free energy, driving the movements of their cilia or flagella.  Many internal degrees of freedom are thus required to describe the configuration of cilia or flagella, in addition to the internal molecular composition of a micro-organism.  

\begin{figure}[h]
\centerline{\scalebox{0.6}{\includegraphics{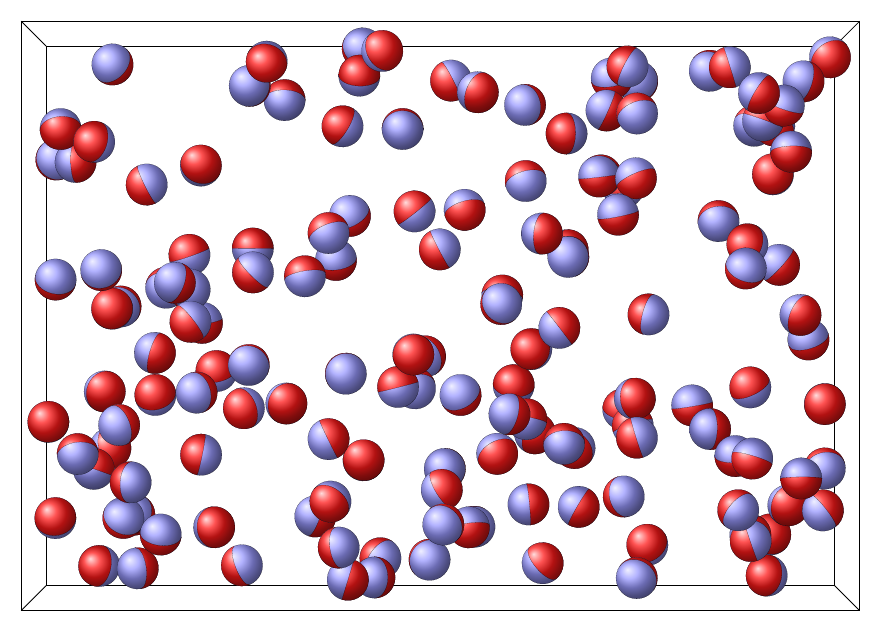}}}
\caption{Illustrative example of an active suspension, which is here composed of Janus particles.  These axisymmetric particles are self-propelled through the surrounding fluid by chemical reactions occurring at their catalytic hemisphere and by diffusiophoresis coupling the concentration gradients of the reactants to the fluid velocity. The case of Janus particles is considered in section~\ref{sec:Janus}.}
\label{fig1}
\end{figure}

In active suspensions, the agents are synthetic colloidal particles of micrometric size immersed in a solution containing reactants, as illustrated in figure~\ref{fig1}.  These particles have the ability to catalyse chemical reactions at their surface and to produce molecular concentration gradients and fluid flows in the neighbouring solution.  For these active particles, diffusiophoresis is the mechanism of coupling between the concentration gradients and the fluid velocity at the surface of each particle, self-generating their propulsion through the surrounding fluid.  The internal degrees of freedom of such rigid colloidal particles are the angles specifying their orientation and, in addition, their possible coverage with adsorbed molecular species.  In such active suspensions, the colloidal particles are thus subjected to hydrodynamic interactions mediated by the velocity field and chemotactic interactions mediated by the molecular concentration fields, in addition to direct mechanical interactions manifesting themselves if the suspension is dense enough.

Therefore, the internal degrees of freedom play key roles in active matter, especially, for the mechanisms of energy transduction, energy dissipation, and entropy production.  Concerning these issues, important efforts have been recently devoted to understand the thermodynamics of active matter \cite{GK17,PS18,JGS18,S19,MFTC21,FJC22,ATR25,BRS25}.  Interestingly, in 1953, Prigogine and Mazur~\cite{PM53} had already extended the thermodynamics of irreversible processes to systems with internal degrees of freedom, formulating the second principle of thermodynamics in the configuration space of the constituent entities, including the subspace of their spatial position and the subspace of their internal degrees of freedom.  The results of their fundamental work were also reported in the textbooks afterwards written by Prigogine~\cite{P55}, and by {de Groot} and Mazur~\cite{GM62}.  More recently, further contributions to the thermodynamics of systems with an internal structure were proposed for modeling the rheological properties of complex fluids \cite{BE94,OG97}.

In active matter, the extension of non-equilibrium thermodynamics to include the internal degrees of freedom is essential to understand the coupling mechanisms that perform the interconversion of free energy into mechanical motion.  Indeed, non-equilibrium thermodynamics provides a unifying theoretical framework to establish the macroscopic equations of motion in consistency with the principles of thermodynamics, the Onsager-Casimir principle of microreversibility, and the Curie principle of symmetry under spatial transformations \cite{P55,GM62,C1894,O31a,O31b,C45,N79}.  In particular, non-equilibrium thermodynamics allows us to derive the entropy production rate from the macroscopic conservation and kinetic laws and to identify the reversible and irreversible couplings in the macroscopic equations ruling the time evolution of the system.  Furthermore, the thermodynamic efficiencies of the energy interconversion mechanisms and their thermodynamic cost can be evaluated with this insightful theory.

In this paper, our purpose is to develop the non-equilibrium thermodynamics of active suspensions by including the internal degrees of freedom specifying the orientation of the colloidal particles into the mathematical description.  In this direction, important contributions were obtained by Brenner and coworkers \cite{B63,B64a,B64b,B65,B67,BC72,HB83}, who calculated the Stokes hydrodynamic friction for particles of arbitrary shape using fluid mechanics and, therefrom, generalised Fick's law to the translational and rotational diffusion processes and the possibility of their coupling for helicoidal (i.e., screw-like) particles.  Moreover, the velocity field in the fluid around the particles is coupled by diffusiophoresis to the molecular concentration fields at the surface of the particles, where chemical reactions occur by heterogeneous catalysis.  In the limit of a dilute suspension, the colloidal particles are far from each other and the fields can be calculated around each individual particles by using the equations of chemohydrodynamics \cite{GK18,GK19,GK20,G25}.  In this way, the propulsion speed of the particles can be obtained in terms of the diffusiophoretic properties and the local reaction rates.  More recently, the methods pioneered by Einstein \cite{E1906,LL59,B70} for the effective viscosity of passive dilute suspensions have been extended to calculate systematically the stress tensor of active suspensions, including the contributions from chemical reactions and molecular concentration gradients \cite{G25}.

Because of the orientation of the colloidal particles, coupling mechanisms may exist in such active suspensions, although they are forbidden by Curie's symmetry principle in simple isotropic solutions with reactions.  By their geometry, the colloidal particles change the consequences of Curie's symmetry principle, allowing the possibility of mechanochemical couplings in active suspensions.

Gathering the previously obtained results and using the Onsager-Casimir reciprocal relations \cite{O31a,O31b,C45}, the non-equilibrium thermodynamics of active suspensions can be completed.  All the non-equilibrium constitutive relations can thus be established using chemohydrodynamics on the basis of the interfacial properties between the solid particles and the fluid solution in terms of the surface reaction rate constants of heterogeneous catalysis, the coefficients characterising diffusiophoresis, the Navier slip length, and the geometry of the colloidal particles.

Hereafter, the paper is organised as follows.  The macroscopic description of active suspensions is set up in section~\ref{sec:description}, defining the macrofields we shall consider.  In section~\ref{sec:equations}, the local conservation and balance equations ruling their time evolution are presented.  The thermodynamics of the suspension is established in section~\ref{sec:thermo}, starting from the local Gibbs and Euler thermodynamic relations, deriving therefrom the local balance equation for entropy, and discussing the possible non-equilibrium constitutive relations between the thermodynamic forces, also called affinities (which drive the suspension away from equilibrium), and the resulting dissipative current densities.  In section~\ref{sec:dilute-suspension}, the formalism is developed for isothermal, incompressible, dilute suspensions, in particular, using the work of Brenner and coworkers \cite{B63,B64a,B64b,B65,B67,BC72,HB83}.  In section~\ref{sec:Janus}, the case of suspensions composed of spherical Janus particles, having a catalytic hemisphere and a non-catalytic one, is studied systematically using the results provided by previous work \cite{GK18,GK19,GK20,G25}.  Conclusion and perspectives are drawn in section~\ref{sec:conclusion}.  The appendices contain supplementary materials on the Eulerian angles, the torque exerted on a rigid body, vector calculus with Eulerian angles, the derivation of the local balance equation for entropy, the equilibrium thermodynamics of dilute suspensions, the connection with the work of Brenner and coworkers, the expansion of the distribution function for axisymmetric particles, the evolution equations for the macrofields of polar and nematic orders, and inequalities resulting from the second law of thermodynamics.

\section{Macroscopic description of active suspensions}
\label{sec:description}

\subsection{Mesostructure and mesodynamics of active suspensions}

An active suspension is composed of catalytic colloidal particles motorized by chemical reactions with molecular species dissolved in the surrounding fluid.  The colloidal particles are rigid bodies of micrometric size and arbitrary shape, which remain invariant in time.  Some portions of their surface are catalytic for reactions with specific solute species.  These reactions induce concentration profiles of these species around each particle, which is thus propelled through the fluid by the so-called diffusiophoretic effect.  Diffusiophoresis arises from the molecular interaction forces between the atoms composing the solid particle and the solute species, and the concentration gradients of these species produced by the catalytic reactions in the surrounding solution.  The local concentration gradients of the reactive molecular species are coupled to the local velocity field along the surface of the particle, generating the propulsion of the particle.  This coupling by self-diffusiophoresis induces the interconversion of chemical free energy into the motion of the colloidal particles in the fluid.  By this mechanism, the particles are self-propelled and the suspension is thus active.  Such micrometric particles can be made of polystyrene or silica spheres with a platinum cap.  Once immersed in a solution with hydrogen peroxide, these particles can reach speeds of the order of 10 $\mu$m/s \cite{VTZKGKO10,KYCS10,CEIG19}.  Other kinds of active particles have been synthesized and experimentally investigated \cite{PKOSACMLC04,WDAMS13,SSK14,GPDW14}.  The colloidal particles may have different geometries.  They may be spherical with a catalytic hemisphere and an inert one, ellipsoidal with a catalytic cap preserving or not their axisymmetry, helicoidal with a screw-like shape, or have the form of a rigid dimer of spheres \cite{GLA07,PDTR10,CRRK14,RK15,RHSK16,NGMS23}.  For simplicity, it is assumed that all the colloidal particles of the suspension are identical and that their surface coverage by adsorbed molecular species, their deeper composition, and, thus, their mass remain time invariant.

The motion of the particles can be determined by solving the equations of fluid mechanics coupled by the surface catalytic reactions and diffusiophoresis to the advection-diffusion equations of the solute molecular species on scales of the order of the size of each particle.  On larger spatial scales, the colloidal particles are statistically distributed in the suspension, which may thus be considered as a continuous medium mathematically described in terms of average macroscopic fields.  Here, our purpose is to obtain the macroscopic equations ruling the time evolution of active suspensions on scales that are larger than the mean inter-particle distance.  On such scales, relevant macrofields can be defined by averaging over the statistical distribution of colloidal particles.

\subsection{The macrofields of the suspension}

At the macroscale, the description of the suspension can be formulated in terms of continuous fields that are the slowest modes of time evolution in the system.  These modes include the locally conserved quantities and the molecular densities that remain out of chemical equilibrium for long enough, because they are involved in relatively slow reactions transforming the molecular species.  The locally conserved quantities include the distribution function of the colloidal particles and the densities of mass, linear momentum, and total energy, assuming the absence of external force field.  The other degrees of freedom are described in terms of the entropy of the suspension.

\subsubsection{The distribution function of the colloidal particles}

Typically, the positions and the orientations of the colloidal particles are statistically distributed in the suspension.  The suspension can thus be described in terms of the average distribution function, $f_{\rm C}({\bf r},\pmb{\alpha},t)$, depending on the position ${\bf r}=(r^1,r^2,r^3)=(x,y,z)\in{\mathbb R}^3$; the three Euler angles $\pmb{\alpha}=(\alpha^1,\alpha^2,\alpha^3)=(\theta,\phi,\psi)$ with $0\le \theta \le\pi$, $0\le \phi,\psi <2\pi$; and the time $t$.  The distribution function has the units $[f_{\rm C}]=$~m$^{-3}$~rad$^{-3}$, where `m' stands for meter and `rad' for radian.
The density of colloidal particles, i.e., their number per unit volume, is defined as $n_{\rm C}({\bf r},t)\equiv \int f({\bf r},\pmb{\alpha},t)\, \ddo$, where $\ddo=\sin\theta\, \dd\theta\, \dd\phi \,\dd\psi$ is the integration element in the space of Euler's angles, such that $\int\ddo=8\pi^2$.  The suspension occupies a three-dimensional spatial domain $\cal V$ of volume $V=\int_{\cal V} \ddv$, $\ddv=\dd^3r=\dd x \, \dd y \, \dd z$ being the volume element of integration.  If there is no particle flux at the boundary of the domain $\cal V$, the number of particles given by $N_{\rm C}(t)\equiv \int_{\cal V} n_{\rm C}({\bf r},t) \, \ddv$ is invariant, $\dd N_{\rm C}/\dd t=0$.  During the time evolution of the system, the colloidal particles keep their shape, their composition, and their mass $m_{\rm C}$, so that their distribution function $f_{\rm C}$ should obey a local conservation equation.

\subsubsection{The densities of molecular species}

The active suspension should also be described in terms of the average densities or concentrations $n_\varkappa({\bf r},t)$ of the molecular species ${\rm X}_\varkappa$ with $\varkappa=0,1,2,\dots,{\mathscr M}$, which are assumed to be point-like, ignoring their stereochemical structure.  These molecules are much smaller and lighter than the colloidal particles, having masses such that $m_\varkappa \ll m_{\rm C}$.  The label $\varkappa=0$ is taken for the solvent species and the labels $\varkappa=1,2,\dots,{\mathscr M}$ for the solute species.  The solvent is usually taken as the most abundant species in the solution.

We emphasize that the molecular densities $n_\varkappa({\bf r},t)$ are defined by averaging over spatial scales larger than the mean inter-particle distance.  The local variations of the molecular concentrations on scales of the order of the particle diameter are thus smoothed out after averaging, as required for our macroscopic description.

The molecular species  ${\rm X}_\varkappa$ and the colloids ${\rm C}$ are involved in the following chemical reactions $r=1,2,\dots,{\mathscr R}$:
\be
\sum_{\varkappa=0}^{\mathscr M} \nu_{\varkappa r}^{(+)} \, {\rm X}_\varkappa + \nu_{{\rm C}r}^{(+)} \, {\rm C} \ \underset{-r}{\stackrel{+r}{\rightleftharpoons}} \ \sum_{\varkappa=0}^{\mathscr M} \nu_{\varkappa r}^{(-)} \, {\rm X}_\varkappa + \nu_{{\rm C}r}^{(-)} \, {\rm C} \, ,
\label{reactions}
\ee
where $\nu_{\varkappa r}^{(\pm)}$ are the numbers of molecules of species $\varkappa$ involved in the forward and backward reactions $\pm r$, respectively.  The stoichiometric coefficient $\nu_{\varkappa r}\equiv \nu_{\varkappa r}^{(-)}-\nu_{\varkappa r}^{(+)}$ is defined as the net number of molecules of species $\varkappa$ produced by the forward reaction $+r$.  The reaction is catalysed by the colloidal particles $\rm C$ if $\nu_{{\rm C}r}^{(+)} =\nu_{{\rm C}r}^{(-)} =1$, or it occurs in the bulk of the solution if $\nu_{{\rm C}r}^{(+)} =\nu_{{\rm C}r}^{(-)} =0$.  Accordingly, the stoichiometric coefficients of the colloids are always equal to zero, $\nu_{{\rm C}r}\equiv \nu_{{\rm C}r}^{(-)} - \nu_{{\rm C}r}^{(+)} =0$, which is in relation to the invariance of the colloids.

The reactions and, in particular, those catalysed by the colloidal particles are assumed to be slower than the time scale to reach chemical equilibrium.  Since the molecules are transformed by the reactions, the molecular densities are ruled by local balance equations.  In the absence of reaction, these equations become local conservation equations for the molecular species.

\subsubsection{The mass density}

The total mass density of the suspension is given by
\be
\rho({\bf r},t) \equiv \sum_{\varkappa = 0}^{\mathscr M} m_\varkappa \, n_\varkappa({\bf r},t) + m_{\rm C} \, n_{\rm C}({\bf r},t) \, ,
\label{mass}
\ee
which is locally conserved by the Lavoisier principle of mass conservation, holding in non-relativistic systems.

\subsubsection{The velocity field}

The velocity field is defined as the velocity of the centre of mass of the suspension element at position~$\bf r$ and time~$t$ on scales larger than the mean inter-particle distance.  Accordingly, the velocity is given by the ratio ${\bf v}({\bf r},t)\equiv{\bf g}({\bf r},t)/\rho({\bf r},t)$ between the average linear momentum density ${\bf g}({\bf r},t)$ and the average mass density~(\ref{mass}).

We suppose that there is no external force field like a gravitational or an electrical field that is exerted on the suspension.  As a consequence, the linear momentum density obeys a local conservation equation, which leads to the Navier-Stokes equations for the velocity field of the suspension.  Moreover, the angular momentum density $\pmb{\ell}\equiv {\bf r}\times{\bf g}$ is also assumed to be locally conserved, so that the rank-two pressure tensor $\pmb{\mathsf P}$ of the suspension, as well as its stress tensor $\pmb{\mathsf\sigma}=-\pmb{\mathsf P}$, are symmetric.

\subsubsection{The total and internal energy densities}

Because of the absence of external force field, the total energy density $\epsilon\equiv e+\rho{\bf v}^2/2$, which is defined as the sum of the average internal energy density $e({\bf r},t)$ and kinetic energy density $\rho{\bf v}^2/2$, is also locally conserved in the suspension.  This local conservation law leads to the heat equation for the temperature.

\section{The local conservation and balance equations}
\label{sec:equations}

The time evolution of the locally conserved quantities and the molecular densities are ruled by partial differential equations, which are here presented.

\subsection{The local conservation equation for the colloidal particles}

The local conservation of the colloidal particles is special because it should be formulated in their six-dimensional configuration space of coordinates ${\bf q}=(q^1,q^2,q^3,q^4,q^5,q^6)=({\bf r},\pmb{\alpha})=(x,y,z,\theta,\phi,\psi)$, giving the position and the orientation of each particle \cite{PM53,B65,B67}.  The position is specified by the three Cartesian coordinates ${\bf r}=(x,y,z)\in{\mathbb R}^3$ and the orientation by the curvilinear coordinates that are the three Eulerian angles $\pmb{\alpha}=(\theta,\phi,\psi)$ for a rotation of the group SO(3), bringing an initial reference orientation to the current orientation of the particle.  

Accordingly, the local conservation equation should be expressed as
\be
\partial_t \, f_{\rm C} + {\rm div} \, {\bf J}_{\rm C} = 0
\qquad\mbox{with}\qquad
{\bf J}_{\rm C} = 
\left(
\begin{array}{c}
{\bf J}_{\rm Ct} \\
{\bf J}_{\rm Cr}
\end{array}
\right)
\label{eq-C}
\ee
in terms of the divergence of the six-dimensional current density ${\bf J}_{\rm C}$, which has three translational components ${\bf J}_{\rm Ct}$ and three rotational components ${\bf J}_{\rm Cr}$.

{\it A priori}, the three-dimensional vector ${\bf J}_{\rm Cr}$ is defined with respect to a Cartesian basis of unit vectors corresponding to rotations of infinitesimal angles $\dd\pmb{\chi}$ taken around the Cartesian axes.  Equivalently, we may consider a basis corresponding to rotations of infinitesimal Eulerian angles $\dd\pmb{\alpha}$.  The two kinds of infinitesimal angles are related to each other according to $\dd\pmb{\chi} =\pmb{\mathsf N}^{\rm T}\cdot \dd\pmb{\alpha}$ with a $3\times 3$ matrix $\pmb{\mathsf N}(\pmb{\alpha})$, which is known in the mechanics of the rigid body \cite{RSGK20} (see appendix~\ref{AppA}).  In the new basis, the three rotational components are given by
\be
\pmb{\mathfrak J}_{\rm Cr} = \pmb{\mathsf N}^{{\rm T}-1} \cdot {\bf J}_{\rm Cr} \, ,
\label{J_Cr}
\ee
where the superscripts denotes the transpose $^{\rm T}$ and the inverse $^{-1}$, respectively (see appendix~\ref{AppB}).  Since the coordinates of position are Cartesian, the three translational components of the current density are unchanged:
\be
\pmb{\mathfrak J}_{\rm Ct} = {\bf J}_{\rm Ct} \, .
\label{J_Ct}
\ee
Therefore, the relation between the Cartesian and curvilinear components of the six-dimensional current density read
\be
\pmb{\mathfrak J}_{\rm C} = 
\left(
\begin{array}{c}
\pmb{\mathfrak J}_{\rm Ct} \\
\pmb{\mathfrak J}_{\rm Cr}
\end{array}
\right)
=\pmb{\mathsf \Theta}^{\rm T}\cdot {\bf J}_{\rm C}
\qquad\mbox{with the $6\times 6$ matrix}\qquad
\pmb{\mathsf \Theta}
\equiv
\left(
\begin{array}{cc}
\pmb{\mathsf I} & 0 \\
0 & \pmb{\mathsf N}^{-1}
\end{array}
\right) ,
\label{J_C-Cartesian-curvilinear}
\ee
where $\pmb{\mathsf I}=(\delta^{ij})$ denotes the $3\times 3$ unit matrix.

In the Cartesian coordinates of position, the contravariant and covariant components of any vector are identical: $X_{\rm t}^i=X_{{\rm t},i}$.  However, since the angular coordinates are curvilinear, the contravariant and covariant components differ.  If $X_{\rm r}^i$ denote the rotational contravariant components, the covariant components are given by $X_{{\rm r},i}=g_{ij}X_{\rm r}^j$ in terms of the $3\times 3$ matrix $(g_{ij})=\pmb{\mathsf g}=\pmb{\mathsf N}\cdot\pmb{\mathsf N}^{\rm T}$, defining the metric of the curvilinear coordinates, $\dd\pmb{\chi}^2 =\pmb{\mathsf g}:\dd\pmb{\alpha}^2 = g_{ij}\, \dd\alpha^i \, \dd\alpha^j$.  Reciprocally, the contravariant components are recovered from the covariant ones according to $X_{\rm r}^i=g^{ij} X_{{\rm r},j}$ with the inverse of this matrix, $(g^{ij})=\pmb{\mathsf g}^{-1}=\pmb{\mathsf N}^{{\rm T}-1}\cdot \pmb{\mathsf N}^{-1}$.  The determinant of the metric is given by $g=\det\pmb{\mathsf g} = \left(\det\pmb{\mathsf N}\right)^2=\sin^2\theta$, which confirms the form, $\ddo=\sqrt{g}\, d^3\alpha=\sin\theta\, \dd\theta \, \dd\phi \, \dd\psi$, for the element of integration in Eulerian angles (see appendix~\ref{AppA}).

In curvilinear coordinates, the local conservation equation~(\ref{eq-C}) reads
\be
\partial_t \, f_{\rm C} + \frac{1}{\sqrt{g}} \, \frac{\partial}{\partial q^i} \left(\sqrt{g} \, {\mathfrak J}_{\rm C}^i\right) = 0
\label{eq-C-2}
\ee
or, equivalently,
\be
\partial_t \, f_{\rm C} +\pmb{\nabla}\cdot\pmb{\mathfrak J}_{\rm Ct} + \frac{1}{\sqrt{g}} \, \frac{\partial}{\partial\pmb{\alpha}} \cdot\left(\sqrt{g} \, \pmb{\mathfrak J}_{\rm Cr}\right) = 0 \, ,
\label{eq-C-3}
\ee
where the divergences in position and orientation spaces are respectively given by
\bea
&&{\rm div}_{\rm t}\, \pmb{\mathfrak J}_{\rm Ct} = \pmb{\nabla}\cdot\pmb{\mathfrak J}_{\rm Ct} = \partial_x\,{\mathfrak J}_{\rm Ct}^x + \partial_y\,{\mathfrak J}_{\rm Ct}^y + \partial_z\,{\mathfrak J}_{\rm Ct}^z \, , \label{dfn-div-t}\\
&&{\rm div}_{\rm r}\, \pmb{\mathfrak J}_{\rm Cr} = \frac{1}{\sqrt{g}} \, \frac{\partial}{\partial\pmb{\alpha}} \cdot\left(\sqrt{g} \, \pmb{\mathfrak J}_{\rm Cr}\right) = \frac{1}{\sin\theta} \, \partial_\theta\left(\sin\theta \, {\mathfrak J}_{\rm Cr}^\theta \right) + \partial_\phi\,{\mathfrak J}_{\rm Cr}^\phi + \partial_\psi\,{\mathfrak J}_{\rm Cr}^\psi \, . \label{dfn-div-r}
\eea

Furthermore, the translational and rotational components of the colloidal current density can split as follows into their reversible part due to advection and another part possibly contributing to dissipation and that we shall refer to as their dissipative part,
\be
\left\{
\begin{array}{c}
\pmb{\mathfrak J}_{\rm Ct} = f_{\rm C} \, {\bf v} + \pmb{\mathscr J}_{\rm Ct} \, , \\
\pmb{\mathfrak J}_{\rm Cr} = f_{\rm C} \, \pmb{\upsilon}_{\rm r} + \pmb{\mathscr J}_{\rm Cr} \, ,
\end{array}
\right.
\label{J_Ct-J_Cr-adv-diss}
\ee
where $\bf v$ is the three-dimensional velocity field of the suspension and
\be
\pmb{\upsilon}_{\rm r}=\pmb{\mathsf N}^{{\rm T}-1}\cdot{\displaystyle\frac{\pmb{\omega}}{2}}
\qquad\mbox{with}\qquad
\pmb{\omega}=\pmb{\nabla}\times{\bf v}
\label{v_r}
\ee
is the rotational velocity, which is known to be given in terms of one half the vorticity $\pmb{\omega}$ of the velocity field \cite{B67,BC72,JGS18,G25}.

Integrating the local conservation equation~(\ref{eq-C-3}) over Euler's angles, we deduce the local conservation equation for the density of colloidal particles, which has the following form,
\be
\partial_t \, n_{\rm C} + \pmb{\nabla}\cdot \pmb{\jmath}_{\rm C} = 0
\qquad\mbox{with the three-dimensional current density}\qquad
\pmb{\jmath}_{\rm C} \equiv \int \pmb{\mathfrak J}_{\rm Ct} \, \ddo = n_{\rm C} \, {\bf v} + \int \pmb{\mathscr J}_{\rm Ct} \, \ddo \, ,
\label{eq-n_C}
\ee
because $\int {\rm div}_{\rm r}\,\pmb{\mathfrak J}_{\rm Cr} \, \ddo =0$.

\subsection{The local balance equations for the molecular species}

Since the molecular species $\varkappa=0,1,2,\dots,{\mathscr M}$ are transported in the suspension and transformed by the ${\mathscr R}$ reactions~(\ref{reactions}), their densities or concentrations $n_\varkappa({\bf r},t)$ obey the following local balance equations,
\be
\partial_t n_\varkappa + \pmb{\nabla}\cdot(n_\varkappa{\bf v}+\pmb{\cal J}_\varkappa) = \sum_{r=1}^{\mathscr R} \nu_{\varkappa r} \, w_r \, , 
\label{eq-n_k}
\ee
where $\pmb{\cal J}_\varkappa$ denotes the dissipative part of the corresponding molecular current density $\pmb{\jmath}_\varkappa = n_\varkappa{\bf v}+\pmb{\cal J}_\varkappa$, $\nu_{\varkappa r}$ is the stoichiometric coefficient of the species $\varkappa$ in the reaction $r$, and $w_r=w_{+r}-w_{-r}$ is the net rate density of the reaction~$r$.

\subsection{The local conservation equation for mass}

The mass density given by equation~(\ref{mass}) should satisfy the continuity equation,
\be
\partial_t \, \rho + \pmb{\nabla}\cdot\left(\rho \, {\bf v}\right) = 0 \, .
\label{eq-mass}
\ee
Since this equation is obtained by combining the local conservation equation~(\ref{eq-n_C}) and the balance equations~(\ref{eq-n_k}) with the masses of the different species, we find the constraint
\be
\sum_{\varkappa=0}^{\mathscr M} m_\varkappa \, \nu_{\varkappa r} = 0 \, ,
\label{constraint-1}
\ee
which is a consequence of the Lavoisier principle of mass conservation in every reaction $r$; and the following other constraint on the dissipative parts of the molecular and colloidal current densities,
\be
\sum_{\varkappa=0}^{\mathscr M} m_\varkappa \, \pmb{\cal J}_\varkappa + m_{\rm C} \int \pmb{\mathscr J}_{\rm Ct} \, \ddo = 0 \, .
\label{constraint-2}
\ee
Therefore, the dissipative current density of one species can be expressed in terms of the other dissipative current densities.  In this regard, the solvent is usually taken as the reference species \cite{JGS18}, so that its dissipative current density can be expressed as
\be
\pmb{\cal J}_0 = - \sum_{\varkappa=1}^{\mathscr M} \frac{m_\varkappa}{m_0} \, \pmb{\cal J}_\varkappa - \frac{m_{\rm C}}{m_0} \int \pmb{\mathscr J}_{\rm Ct} \, \ddo \, .
\label{J_0}
\ee

\subsection{The local conservation equation for linear momentum}

We assume that there is no external force and no external torque exerted on the suspension.  Accordingly, the average linear momentum density ${\bf g}=\rho{\bf v}$ obeys the following local conservation law,
\be
\partial_t(\rho{\bf v})+\pmb{\nabla}\cdot(\rho{\bf v}{\bf v} + \pmb{\mathsf P}) = 0 \, , 
\qquad\mbox{where}\qquad
\pmb{\mathsf P} = p \, \pmb{\mathsf I} + \pmb{\mathsf \Pi}
\label{eq-g}
\ee
is the rank-two pressure tensor composed of the (hydrostatic) pressure $p$ multiplied by the $3\times 3$ unit tensor $\pmb{\mathsf I}$, and its dissipative part $\pmb{\mathsf \Pi}$. The latter contains the contributions from shear and bulk viscosities, diffusiophoresis, and self-diffusiophoresis due to the presence and the reactions of molecular and colloidal species in the suspension \cite{R10,G25}.  Moreover, this tensor is symmetric because of the local conversation of angular momentum: $\pmb{\mathsf \Pi}=\pmb{\mathsf \Pi}^{\rm T}$.

\subsection{The local conservation equation for the total energy}

In the absence of an external force field, the total energy density $\epsilon= e+\rho{\bf v}^2/2$ satisfies the following local conservation equation,
\be
\partial_t\epsilon + \pmb{\nabla}\cdot(\epsilon{\bf v} + \pmb{\mathsf P}\cdot{\bf v} +\pmb{\cal J}_q) = 0 \, , 
\label{eq-etot}
\ee
where the term $\pmb{\mathsf P}\cdot{\bf v}$ gives the contribution of the work performed by the pressure tensor and $\pmb{\cal J}_q$ is the heat current density, which is dissipative.  Therefrom, the local balance equation for the internal energy density $e$ can be derived and, consequently, the heat equation for the local temperature.

\vskip 0.3 cm

The local conservation and balance equations (\ref{eq-C-3}), (\ref{eq-n_k}), (\ref{eq-mass}), (\ref{eq-g}), and (\ref{eq-etot}) rule the time evolution of the slowest modes of the suspension, i.e., the modes which are observable at the macroscale.  However, these equations are not yet complete because the dissipative current densities and the reaction rate densities are still unknown at this stage of the formulation.

\section{The thermodynamics of the suspension}
\label{sec:thermo}

In order to obtain a closed set of equations for the macrofields, we need constitutive relations for the dissipative current densities and the reaction rate densities.  For this purpose, we use the principles of non-equilibrium thermodynamics \cite{P55,GM62,N79}.  In this way, constitutive relations can be established, which are consistent with the second law of thermodynamics, microreversibility, and the symmetries of the system under spatial transformations, i.e., Curie's symmetry principle.  To formulate the non-equilibrium thermodynamics of active suspensions, we need to first introduce the entropy, which is the thermodynamic potential of the system and, next, to derive the balance equation for the entropy density using the local Gibbs and Euler thermodynamic relations and the local conservation and balance equations of the macrofields, which have been given in the previous section~\ref{sec:equations}.

\subsection{The entropy of the suspension}

The suspension is assumed to be characterised by an entropy functional, $S=S[e,\{ n_\varkappa \},f_{\rm C},$ $\pmb{\nabla} e,\{ \pmb{\nabla} n_\varkappa \},\pmb{\nabla} f_{\rm C},$ $\partial_{\pmb{\alpha}}f_{\rm C},\dots]$, which depends on the internal energy density, the molecular densities, the colloidal distribution function, and, possibly, their gradients.  Moreover, this entropy is supposed to be given by the volume integral of an entropy density, which is also depending on these macrofields: $S=\int_{\cal V} s({\bf r},t) \, \ddv$.  The entropy density represents a thermodynamic potential defined in a frame moving with each element of the suspension.  The entropy accounts for the fast degrees of freedom of the suspension that locally equilibrate more quickly than the slow macroscopic modes of the description.  The dependence on the gradients is required in heterogeneous systems, having spatial scales between the size of the colloidal particles and the macroscale, and which are described for instance by Ginzburg-Landau or Cahn-Hilliard functionals \cite{GL50,CH58,PF90}.  Otherwise, the entropy may be assumed to depend only on the macrofields themselves as $S=S[e,\{ n_\varkappa \},f_{\rm C}]$.

For thermodynamic consistency, the entropy density should satisfy the following local Gibbs differential relation:
\be
\delta s = \frac{1}{T} \, \delta e - \sum_{\varkappa = 0}^{\mathscr M} \frac{\mu_\varkappa}{T} \, \delta n_\varkappa - \int \ddo \, \frac{\zeta_{\rm C}}{T} \, \delta f_{\rm C} \, ,
\label{Gibbs}
\ee
where $T({\bf r},t)$ is the local temperature, $\mu_\varkappa({\bf r},t)$ is the local chemical potential of the molecular species $\varkappa$, and $\zeta_{\rm C}({\bf r},\pmb{\alpha},t)$ is the chemical potential in the configuration space of the colloidal particles \cite{PM53,P55,GM62}.  The latter is obtained by taking the functional derivative of the entropy with respect to some variation $\delta f_{\rm C}$ of the colloidal distribution function.  Moreover, the entropy density should also obey the local Euler relation given by
\be
s = \frac{1}{T} \left( e + p \right)  - \sum_{\varkappa = 0}^{\mathscr M} \frac{\mu_\varkappa}{T} \, n_\varkappa - \int \ddo \, \frac{\zeta_{\rm C}}{T} \, f_{\rm C} \, ,
\label{Euler}
\ee
where $p({\bf r},t)$ is the local pressure in the suspension.  The local Gibbs and Euler relations~(\ref{Gibbs}) and~(\ref{Euler}) hold in the frame moving with the element at position $\bf r$ and time $t$ in the suspension.  The entropy density provides the local equilibrium thermodynamic properties of the suspension.  In particular, the equations of state for the pressure and the internal energy can be deduced therefrom in terms of the local temperature and composition of the suspension.  Similar assumptions of local thermodynamic equilibrium are considered for active gels or active nematics \cite{JGS18}.

\subsection{The local balance equation for entropy}

Using the local Gibbs and Euler relations (\ref{Gibbs}) and (\ref{Euler}), the entropy density $s$ can be related to the densities of internal energy, molecular species, and the distribution function of the colloidal particles, which are ruled by the local conservation and balance equations presented in section~\ref{sec:equations}.  Taking moreover into account the constraints (\ref{constraint-1}) and~(\ref{constraint-2}) due to mass conservation, we can derive the following local balance equation for the entropy,
\be
\partial_t s + \pmb{\nabla}\cdot(s\,{\bf v}+\pmb{\cal J}_s) = \sigma_s \ge 0 \, , 
\label{eq-s}
\ee
where the dissipative part of the entropy current density is given by
\be
\pmb{\cal J}_s = \frac{1}{T} \, \pmb{\cal J}_q - \sum_{\varkappa=1}^{\mathscr M} \frac{\overline{\mu}_\varkappa}{T} \,  \pmb{\cal J}_\varkappa - \int \frac{\overline{\zeta}_{\rm C}}{T} \, \pmb{\mathscr J}_{\rm Ct} \, \ddo
\label{J_s}
\ee
in terms of the relative chemical potentials
\bea
&&\overline{\mu}_\varkappa \equiv \mu_\varkappa - \frac{m_\varkappa}{m_0}\, \mu_0 \, , \label{rel-mu_k} \\
&&\overline{\zeta}_{\rm C} \equiv \zeta_{\rm C} - \frac{m_{\rm C}}{m_0}\, \mu_0 \, , \label{rel-zeta_C}
\eea
which are here defined like in Ref.~\cite{JGS18}; and the entropy production rate density $\sigma_s$ can be expressed as
\bea
\frac{1}{k_{\rm B}} \, \sigma_s = \sum_{r=1}^{\mathscr R} {\cal A}_r \, w_r + \sum_{\varkappa=1}^{\mathscr M} \pmb{\cal A}_\varkappa \cdot \pmb{\cal J}_\varkappa + \int \pmb{\mathscr A}_{\rm Ct} \cdot \pmb{\mathscr J}_{\rm Ct} \, \ddo + \int \pmb{\mathscr A}_{\rm Cr} \cdot \pmb{\mathscr J}_{\rm Cr} \, \ddo \, + \, \pmb{\cal A}_q \cdot \pmb{J}_q \, + \stackrel{\circ}{{\pmb{\mathsf A}}}_{\bf g}\, : \, \stackrel{\circ}{{\pmb{\mathsf \Pi}}} + {\cal A}_{\bf g} \, \Pi \ge 0
\label{sigma_s}
\eea
in terms of the affinities and dissipative current densities presented in table~\ref{Tab1}.  The affinities are the thermodynamic forces of the processes driving the system away from equilibrium \cite{P55,GM62,N79}.  They are here defined by dividing their usual expressions with Boltzmann's constant $k_{\rm B}$ to let appear the thermal energy $k_{\rm B}T$ instead of the temperature $T$.  In particular, the affinity of the chemical reaction $r$ is dimensionless and given by ${\cal A}_r=-\Delta G_r/(k_{\rm B}T)$ in terms of the Gibbs free energy of the reaction, $\Delta G_r=\sum_{\varkappa=0}^{\mathscr M} \mu_\varkappa \, \nu_{\varkappa r}=\sum_{\varkappa=1}^{\mathscr M} \overline{\mu}_\varkappa \, \nu_{\varkappa r}$.
The calculation of the expressions~(\ref{J_s}) and~(\ref{sigma_s}) is carried out in appendix~\ref{AppC}.

\begin{table}
\caption{The affinities ${\cal A}_a$ and dissipative current densities ${\cal J}_a$ of the different processes possibly running in active suspensions.  The column `time' gives the parity $\epsilon_a$ of the affinity under time reversal and the column `space' its tensorial character. $k_{\rm B}$~denotes Boltzmann's constant, $w_r$ the rate density of reaction $r$, $T$ the temperature, $\overline{\mu}_\varkappa$ the relative chemical potential (\ref{rel-mu_k}) of molecular species $\varkappa$, $\nu_{\varkappa r}$ its stoichiometric coefficient in reaction $r$, $\overline{\zeta}_{\rm C}$ the colloidal relative chemical potential~(\ref{rel-zeta_C}), ${\stackrel{\circ}{\pmb{\mathsf\Pi}}}\equiv\pmb{\mathsf \Pi}-\Pi\, \pmb{\mathsf I}$ the traceless part of the dissipative pressure tensor $\pmb{\mathsf \Pi}$, $\Pi\equiv({\rm tr}\,\pmb{\mathsf\Pi})/3$ one third of its trace, and $\bf v$ the velocity field.  We note that the dissipative current density of rotational colloidal transport has contravariant components $\pmb{\mathscr J}_{\rm Cr}=({\mathscr J}_{\rm Cr}^{i})$ and the corresponding affinity has covariant components $\pmb{\mathscr A}_{\rm Cr}=({\mathscr A}_{{\rm Cr},i})$, because the derivatives with respect to Euler's angles are covariant $\partial_{\pmb{\alpha}}=(\partial_{\alpha^i})$.  The other vectors and tensors have equal contravariant and covariant components because they are defined in Cartesian spatial coordinates.}
\label{Tab1}
\vskip 0.1 cm
\begin{tabular}{lllcc}
\hline\hline
process & \shortstack{dissipative current \\ density ${\cal J}_a$} & affinity ${\cal A}_a$ & time $\epsilon_a$ & space\\
\hline\\[-8pt]
reaction $r$ & $w_r$ & $\displaystyle {\cal A}_r \equiv - \frac{1}{k_{\rm B}T} \sum_{\varkappa=1}^{\mathscr M} \overline{\mu}_\varkappa \, \nu_{\varkappa r}$ & $+1$ & scalar \\
\shortstack{molecular transport of \\ solute species $\varkappa$} & $\pmb{\cal J}_\varkappa$ & $\displaystyle \pmb{\cal A}_\varkappa \equiv \pmb{\nabla}\left(-\frac{\overline{\mu}_\varkappa}{k_{\rm B}T}\right)$ & $+1$ & vector \\[+10pt]
translational colloidal transport \quad & $\pmb{\mathscr J}_{\rm Ct}$ & $\displaystyle \pmb{\mathscr A}_{\rm Ct} \equiv  \pmb{\nabla}\left(-\frac{\overline{\zeta}_{\rm C}}{k_{\rm B}T}\right)$ & $+1$ & \shortstack{$\pmb{\alpha}$-distributed\\ vector} \\[+10pt]
rotational colloidal transport & $\pmb{\mathscr J}_{\rm Cr}$ & $\displaystyle \pmb{\mathscr A}_{\rm Cr} \equiv  \frac{\partial}{\partial\pmb{\alpha}}\left(-\frac{\overline{\zeta}_{\rm C}}{k_{\rm B}T}\right)$ & +1 & \shortstack{$\pmb{\alpha}$-distributed \\ angular vector} \\[+10pt]
heat conduction & $\pmb{\cal J}_q$ & $\displaystyle \pmb{\cal A}_q \equiv \pmb{\nabla}\left(\frac{1}{k_{\rm B}T}\right)$ & $+1$ & vector \\[+7pt]
\shortstack{linear momentum transport\\ in shear flows} & ${\stackrel{\circ}{\pmb{\mathsf\Pi}}}$ & $\displaystyle \stackrel{\circ}{{\pmb{\mathsf A}}}_{\bf g} \, \equiv -\frac{1}{2k_{\rm B} T} \left(\pmb{\nabla}{\bf v}+\pmb{\nabla}{\bf v}^{\rm T}-\frac{2}{3}\, \pmb{\nabla}\cdot{\bf v}\, \pmb{\mathsf I}\right)$ & $-1$ &tensor \\[+7pt]
\shortstack{linear momentum transport\\ in dilatational flows} & $\Pi$ &  $\displaystyle {\cal A}_{\bf g}=-\frac{1}{k_{\rm B}T} \, \pmb{\nabla}\cdot {\bf v}$ & $-1$ & scalar \\[+10pt]
\hline\hline
\end{tabular}
\end{table}

The entropy production rate density~(\ref{sigma_s}) should always be non-negative to satisfy the second law of thermodynamics.  Its calculation shows that it has the generic form of a sum over the irreversible processes of their affinity multiplied by their dissipative current density:
\be
\frac{1}{k_{\rm B}} \, \sigma_s = \sum_a {\cal A}_a \, {\cal J}_a \ge 0
\label{sigma_s-generic}
\ee
(which is here combined with angular integrals for the $\pmb{\alpha}$-distributed quantities).
Accordingly, the time derivative of the entropy $S=\int_{\cal V} s \, \ddv$ is given by
\be
\frac{\dd S}{\dd t} = \frac{\dd_{\rm e} S}{\dd t} + \frac{\dd_{\rm i} S}{\dd t} 
\label{dSdt}
\ee
in terms of the rate of entropy exchange with the environment of the system
\be
\frac{\dd_{\rm e} S}{\dd t} = - \oint_{\partial{\cal V}} (s\, {\bf v} + \pmb{\cal J}_s) \cdot \dd\pmb{\Sigma}
\label{deSdt}
\ee
and the rate of entropy production inside the system
\be
\frac{\dd_{\rm i} S}{\dd t} = \int_{\cal V} \sigma_s \, \ddv = k_{\rm B} \sum_a \int_{\cal V} {\cal A}_a \, {\cal J}_a \, \ddv \ge 0 \, ,
\label{diSdt}
\ee
which is non-negative in agreement with the second law of thermodynamics.
For such spatially extended systems, the thermodynamic efficiency of energy transduction of the processes $b(\ne a)$ into the process $a$ can thus be defined as
\be
\eta_a = - \frac{\int_{\cal V} {\cal A}_a \, {\cal J}_a \, \ddv}{\sum_{b(\ne a)} \int_{\cal V} {\cal A}_b \, {\cal J}_b \, \ddv} \leq 1 \, ,
\label{eff_a}
\ee
which is always lower or equal to the unit value because of the second law \cite{H77,G22}.

\subsection{The non-equilibrium constitutive relations}

The key issue is to establish the non-equilibrium constitutive relations between the dissipative current densities $\{ {\cal J}_a \}$ and the affinities $\{ {\cal A}_a \}$, which are allowed by the spatiotemporal symmetries of the system in accordance with Curie's symmetry principle and microreversibility.  These constitutive relations should determine all the possible couplings generating energy interconversion between the processes.  In active matter, these couplings are essential for the transduction of chemical free energy into motion.  Since the affinities and the dissipative current densities should be equal to zero at thermodynamic equilibrium, the latter can be expanded as
\be
{\cal J}_a = \sum_b {\cal L}_{a,b} \, {\cal A}_b + \frac{1}{2} \sum_{b,c} {\cal M}_{a,bc} \, {\cal A}_b \, {\cal A}_c + \cdots
\label{J_a-A_a}
\ee
in terms of linear and nonlinear response coefficients ${\cal L}_{a,b}$, ${\cal M}_{a,bc}$, $\dots$.  According to non-equilibrium statistical mechanics, the linear response coefficients ${\cal L}_{a,b}$ are given by Green-Kubo formulas if the temporal correlations of the microscopic currents decay fast enough with respect to the time scales of the macrofields \cite{G22}.  Otherwise, time-dependent memory effects may arise like in viscoelasticity \cite{DM65}.  The linear approximation, which consists in neglecting the effects of the nonlinear response coefficients in the expansion~(\ref{J_a-A_a}), holds if the macrofields vary over spatial scales that are significantly larger than the mean free paths of the molecular and colloidal species in the system, which is usually the case for heat conduction, diffusion, and flow processes.  Similar conditions concern the angular distribution of colloids.

However, such conditions do not apply to reactions, because their free energy landscape varies over spatial scales of the same order of magnitude as the molecular sizes and, for this reason, their affinity is given by a difference of free energy and not by a gradient.  Therefore, the constitutive relation between their rate density $w_r$ and the corresponding chemical affinity ${\cal A}_r$ is typically nonlinear, unless the system is very close to chemical equilibrium.  Indeed, according to the mass-action law holding in dilute solutions \cite{GM62}, the rate densities of the forward and backward reactions~(\ref{reactions}) are proportional to the densities of molecules or particles involved in the reaction,
\be
w_{\pm r} = k_{\pm r} \prod_{\varkappa=1}^{\mathscr M} n_\varkappa^{\nu_{\varkappa r}^{(\pm)}} \, n_{\rm C}^{\nu_{{\rm C}r}^{(\pm)}} \, ,
\ee
where $k_{\pm r}$ are rate constants and if the solvent is assumed to be inert.  Since the chemical potentials of the solute molecular species are given by $\mu_\varkappa=\mu_\varkappa^0 + k_{\rm B}T \ln(n_\varkappa/n_0)$ in dilute solutions (see appendix~\ref{AppD}) and $\nu_{{\rm C}r}^{(+)}=\nu_{{\rm C}r}^{(-)}$, the net reaction rate density can be written as
\be
w_r = w_{+r}-w_{-r} = w_{-r} \left( {\rm e}^{{\cal A}_r}-1 \right) = w_{-r} \left({\cal A}_r + \frac{1}{2}\, {\cal A}_r^2 + \cdots \right) 
\label{w_r-A_r} \, ,
\ee
which is a strongly nonlinear constitutive relation with the following linear and nonlinear response coefficients,
\be
{\cal L}_{r,r'} = w_{-r} \, \delta_{rr'} \, , \qquad
{\cal M}_{r,r'r''} = w_{-r} \, \delta_{rr'} \, \delta_{rr''} \, , \qquad
\dots
\label{L_rr-M_rrr}
\ee

The principle of microreversibility implies that the linear response coefficients should satisfy the Onsager-Casimir reciprocal relations
\be
{\cal L}_{a,b} = \epsilon_a \, \epsilon_b \, {\cal L}_{b,a} \, ,
\label{Onsager-Casimir}
\ee
where $\epsilon_a=\pm 1$ (resp. $\epsilon_b=\pm 1$) is the parity under time reversal of the corresponding affinity ${\cal A}_a$ (resp. ${\cal A}_b$) \cite{O31a,O31b,C45}.  These parities are given in table~\ref{Tab1} for each process.  Since the coefficient ${\cal L}_{a,b}$ for $a\ne b$ characterises the coupling between the processes $a$ and $b$, there should exist a reciprocal coupling $b\to a$ of equal coefficient associated with every coupling $a\to b$, because of the relation~(\ref{Onsager-Casimir}).  Microreversibility also implies some constraints on the nonlinear response coefficients \cite{G22}.  

As a consequence of the expansion~(\ref{J_a-A_a}), the entropy production rate density reads
\be
\frac{1}{k_{\rm B}} \, \sigma_s = \sum_{a,b} {\cal L}_{a,b}^{\rm S} \, {\cal A}_a \, {\cal A}_b + O(\pmb{\cal A}^3) \ge 0 \, ,
\label{sigma_s-lin}
\ee
where ${\cal L}_{a,b}^{\rm S}\equiv ({\cal L}_{a,b}+{\cal L}_{b,a})/2$ are the symmetric parts of the linear response coefficients.  Because of the Onsager-Casimir reciprocal relations, only the couplings ${\cal L}_{a,b}$ between processes with equal parities ($\epsilon_a=\epsilon_b$) under time reversal do contribute to entropy production.  Nevertheless, the efficiencies~(\ref{eff_a}) may depend on both the symmetric and antisymmetric parts of the linear response coefficients.  The couplings between processes with opposite parities ($\epsilon_a\ne\epsilon_b$) may thus influence the efficiencies of energy transduction mechanisms.

In these regards, the response coefficients determine the possible couplings in active matter, in particular, between the reactions and the other processes since the activity is powered by the chemical free energy supplied by the reactions.  If there was no coupling to the other processes, the chemical free energy would be lost by direct dissipation within the system.

The suspension is a fluid, but the colloidal particles have in general anisotropic properties, so that their orientation plays a key role.  Therefore, there may exist couplings between the reactions, which have scalar properties, and the other processes, which are vectorial or tensorial, while remaining consistent with Curie's symmetry principle.  However, such couplings would be impossible in a system containing only reactive molecules with point-like behaviour.  This is an essential aspect that makes active suspensions different from usual reactive solutions.

Now, the issue is to obtain the response coefficients of the non-equilibrium constitutive relations in active suspensions.

\section{Isothermal, incompressible, dilute suspensions}
\label{sec:dilute-suspension}

\subsection{The non-equilibrium constitutive relations}
\label{sec:dilute-suspension-A}

To be specific, let us consider active suspensions powered by chemical reactions that are neither exothermic, nor endothermic.  Under such circumstances, the system may be assumed to be isothermal, so that the heat current density and the corresponding affinity are equal to zero, $\pmb{\cal J}_q=0$ and $\pmb{\cal A}_q=0$.

Moreover, a suspension is often a liquid that may be supposed to be incompressible, $\pmb{\nabla}\cdot{\bf v}=0$, which implies that the mass density is uniform and invariant.  Accordingly, the trace of the dissipative part of the pressure tensor and the corresponding affinity are also equal to zero, $\Pi=0$ and ${\cal A}_{\bf g}=0$.  Consequently, the dissipative current density and the affinity associated with linear momentum transport in shear flows reduce to ${\stackrel{\circ}{\pmb{\mathsf\Pi}}}=\pmb{\mathsf\Pi}$ and $\stackrel{\circ}{\pmb{\mathsf A}}_{\bf g}\, =\pmb{\mathsf A}_{\bf g}=-(\pmb{\nabla}{\bf v}+ \pmb{\nabla}{\bf v}^{\rm T})/(2k_{\rm B}T)$, respectively.

In such an active suspension, the problem is to determine the linear response coefficients for the reactions, the molecular and colloidal transports, and the linear momentum transport in shear flows.  For the reactions, we assume that the system is close to chemical equilibrium, so that the reaction rates are linearly related to the corresponding affinities.  Curie's symmetry principle is used to infer the most general tensorial character of the coupling coefficients, which might {\it a priori} be possible given the tensorial characters of table~\ref{Tab1} for the dissipative current density and the affinity that are coupled together by each coefficient.  With the convention of summation over repeated indices, the constitutive relations thus read
\bea
&& w_r = \sum_{r'} {\cal L}_{r,r'} \, {\cal A}_{r'} + \sum_{\varkappa} {\cal L}_{r,\varkappa}^i \, {\cal A}_\varkappa^i + \int {\cal L}_{r,{\rm Ct}}^i \, {\mathscr A}_{\rm Ct}^i \, \ddo + \int {\cal L}_{r,{\rm Cr}}^i \, {\mathscr A}_{{\rm Cr},i} \, \ddo + {\cal L}_{r,{\bf g}}^{ij} \, {\mathsf A}_{\bf g}^{ij} \, , \label{w_r-L}\\
&& {\cal J}_\varkappa^i = \sum_{r} {\cal L}_{\varkappa,r}^i \, {\cal A}_{r} + \sum_{\varkappa'} {\cal L}_{\varkappa,\varkappa'}^{ij} \, {\cal A}_{\varkappa'}^j + \int {\cal L}_{\varkappa,{\rm Ct}}^{ij} \, {\mathscr A}_{\rm Ct}^j \, \ddo + \int {\cal L}_{\varkappa,{\rm Cr}}^{ij} \, {\mathscr A}_{{\rm Cr},j}\, \ddo + {\cal L}_{\varkappa,{\bf g}}^{ijk} \, {\mathsf A}_{\bf g}^{jk} \, , \label{J_k-L}\\
&& {\mathscr J}_{\rm Ct}^i = \sum_{r} {\cal L}_{{\rm Ct},r}^i \, {\cal A}_{r} + \sum_{\varkappa} {\cal L}_{{\rm Ct},\varkappa}^{ij} \, {\cal A}_{\varkappa}^j + \int {\mathscr L}_{{\rm Ct},{\rm Ct}}^{ij} \, {\mathscr A}_{\rm Ct}^j \, \ddo + \int {\mathscr L}_{{\rm Ct},{\rm Cr}}^{ij} \, {\mathscr A}_{{\rm Cr},j}\, \ddo + {\cal L}_{{\rm Ct},{\bf g}}^{ijk} \, {\mathsf A}_{\bf g}^{jk} \, , \label{J_Ct-L}\\
&& {\mathscr J}_{\rm Cr}^i = \sum_{r} {\cal L}_{{\rm Cr},r}^i \, {\cal A}_{r} + \sum_{\varkappa} {\cal L}_{{\rm Cr},\varkappa}^{ij} \, {\cal A}_{\varkappa}^j + \int {\mathscr L}_{{\rm Cr},{\rm Ct}}^{ij} \, {\mathscr A}_{\rm Ct}^j \, \ddo + \int {\mathscr L}_{{\rm Cr},{\rm Cr}}^{ij} \, {\mathscr A}_{{\rm Cr},j}\, \ddo + {\cal L}_{{\rm Cr},{\bf g}}^{ijk} \, {\mathsf A}_{\bf g}^{jk} \, , \label{J_Cr-L}\\
&& \Pi^{ij} = \sum_{r} {\cal L}_{{\bf g},r}^{ij} \, {\cal A}_{r} + \sum_{\varkappa} {\cal L}_{{\bf g},\varkappa}^{ijk} \, {\cal A}_{\varkappa}^k + \int {\cal L}_{{\bf g},{\rm Ct}}^{ijk} \, {\mathscr A}_{\rm Ct}^k \, \ddo + \int {\cal L}_{{\bf g},{\rm Cr}}^{ijk} \, {\mathscr A}_{{\rm Cr},k}\, \ddo + {\cal L}_{{\bf g},{\bf g}}^{ijkl} \, {\mathsf A}_{\bf g}^{kl} \, , \label{Pi-L}
\eea
where the dissipative current densities are contravariant and the affinities are in general covariant.  The latter are equal to their contravariant form in the cases where they are associated with transport in position space and expressed in Cartesian coordinates.

\vskip 0.2 cm

Because of microreversibility, the linear response coefficients should satisfy the following Onsager reciprocal relations (${\cal L}_{a,b}={\cal L}_{b,a}$) between processes with identical parities $\epsilon_a=\epsilon_b$ under time reversal:
\bea
&& {\cal L}_{r,r'}={\cal L}_{r',r} \; , \qquad {\cal L}_{\varkappa,\varkappa'}^{ij}={\cal L}_{\varkappa',\varkappa}^{ji} \; , \qquad {\cal L}_{{\bf g},{\bf g}}^{ijkl}={\cal L}_{{\bf g},{\bf g}}^{klij} \; , \label{ORR_std}\\
&& {\cal L}_{r,\varkappa}^i = {\cal L}_{\varkappa,r}^i \; , \qquad {\cal L}_{r,{\rm Ct}}^i = {\cal L}_{{\rm Ct},r}^i \; , \qquad {\cal L}_{r,{\rm Cr}}^i = {\cal L}_{{\rm Cr},r}^i \; , \label{ORR_r}\\
&& {\cal L}_{\varkappa,{\rm Ct}}^{ij} = {\cal L}_{{\rm Ct},\varkappa}^{ji} \; , \qquad {\cal L}_{\varkappa,{\rm Cr}}^{ij} = {\cal L}_{{\rm Cr},\varkappa}^{ji} \; , \label{ORR_k}\\
&& {\mathscr L}_{{\rm Ct},{\rm Ct}}^{ij} ={\mathscr L}_{{\rm Ct},{\rm Ct}}^{ji} \; , \qquad
{\mathscr L}_{{\rm Ct},{\rm Cr}}^{ij} ={\mathscr L}_{{\rm Cr},{\rm Ct}}^{ji} \; , \qquad
{\mathscr L}_{{\rm Cr},{\rm Cr}}^{ij} ={\mathscr L}_{{\rm Cr},{\rm Cr}}^{ji} \; .
\label{ORR_C}
\eea

The coefficients in the relations~(\ref{ORR_std}) are those for the standard kinetic and transport properties. ${\cal L}_{r,r'}$ is given by the first expression in equation~(\ref{L_rr-M_rrr}).  
${\cal L}_{\varkappa,\varkappa'}^{ij}$ leads to the coefficients of diffusion $D_\varkappa$ and possible cross-diffusion $D_{\varkappa,\varkappa'}$ for the solute molecular species.  As shown in appendix~\ref{AppD}, the solute molecular species have the chemical potentials $\mu_\varkappa=\mu_\varkappa^0 + k_{\rm B}T \ln(n_\varkappa/n_0)$ in dilute solutions.  Therefore, the affinities of molecular transport are given by $\pmb{\cal A}_\varkappa = -\pmb{\nabla}n_\varkappa/n_\varkappa$, because $\Vert\pmb{\nabla}n_\varkappa/n_\varkappa\Vert \gg \Vert\pmb{\nabla}n_0/n_0\Vert$ in a dilute solution where $n_0 \gg n_\varkappa$.  Since the solution is a fluid, which has isotropic properties, the linear response coefficients are related to the diffusion coefficients by ${\cal L}_{\varkappa,\varkappa'}^{ij}=n_{\varkappa'} D_{\varkappa,\varkappa'} \delta^{ij}$.  If cross-diffusion is negligible, we have that $D_{\varkappa,\varkappa'}=D_{\varkappa}\delta_{\varkappa\varkappa'}$.  Furthermore, the rank-four tensor ${\cal L}_{{\bf g},{\bf g}}^{ijkl}$ is proportional to the shear viscosity $\eta$ of the suspension according to ${\cal L}_{{\bf g},{\bf g}}^{ijkl}=k_{\rm B}T \, \eta \left( \delta^{ik}\delta^{jl}+\delta^{il}\delta^{jk} -\frac{2}{3} \delta^{ij}\delta^{kl}\right)$.  The coefficients~(\ref{ORR_r}) characterise the couplings between the chemical reaction $r$ and the molecular and colloidal species due to diffusiophoresis.  The coefficients~(\ref{ORR_k}) give the couplings between the gradients of molecular and colloidal chemical potentials due to diffusiophoresis.  The coefficients~(\ref{ORR_C}) concern the translational and rotational diffusive transports of the colloidal particles.  The coupling coefficients ${\mathscr L}_{{\rm Ct},{\rm Cr}}^{ij}$ between translation and rotational colloidal transports arise if the particles are helicoidal (i.e., screw-like) \cite{B64a,B65}.  We note that the terms associated with the coefficients~(\ref{ORR_std})-(\ref{ORR_C}) contribute to the entropy production rate density~(\ref{sigma_s-lin}) and these terms are thus irreversible in the macroscopic equations of motion (\ref{eq-C-3}), (\ref{eq-n_k}), (\ref{eq-g}), and~(\ref{eq-etot}).

\vskip 0.2 cm

In addition, there are coefficients coupling together processes having different parities $\epsilon_a\ne \epsilon_b$ under time reversal, which obey Onsager-Casimir reciprocal relations (${\cal L}_{a,b}=-{\cal L}_{b,a}$):
\bea
&& {\cal L}_{r,{\bf g}}^{ij} = - {\cal L}_{{\bf g},r}^{ji} \; , \label{CORR-r-g}\\
&& {\cal L}_{\varkappa,{\bf g}}^{ijk} = - {\cal L}_{{\bf g},\varkappa}^{jki} \; , \label{CORR-k-g}\\
&& {\cal L}_{{\rm Ct},{\bf g}}^{ijk} = - {\cal L}_{{\bf g},{\rm Ct}}^{jki} \; , \qquad
{\cal L}_{{\rm Cr},{\bf g}}^{ijk} = - {\cal L}_{{\bf g},{\rm Cr}}^{jki} \; . \label{CORR-C-g}
\eea
These coefficients couple together linear momentum transport to reaction in equation~(\ref{CORR-r-g}), to molecular transport in equation~(\ref{CORR-k-g}), and to colloidal transport in equation~(\ref{CORR-C-g}).  Contrary to the previous coefficients, the terms with the coefficients~(\ref{CORR-r-g})-(\ref{CORR-C-g}) do not contribute to the entropy production rate density~(\ref{sigma_s-lin}) and they are thus reversible in the macroscopic equations~(\ref{eq-C-3}), (\ref{eq-n_k}), (\ref{eq-g}), and~(\ref{eq-etot}).
Moreover, since the dissipative part of the pressure tensor is symmetric $\Pi^{ij}=\Pi^{ji}$ and traceless $\Pi^{ii}=0$, the coefficients~(\ref{CORR-r-g})-(\ref{CORR-C-g}) should also satisfy the following relations:
\bea
&& {\cal L}_{{\bf g},r}^{ij} = {\cal L}_{{\bf g},r}^{ji} \; , \qquad {\cal L}_{{\bf g},r}^{ii} = 0 \; ,\label{sym-tr0-r-g}\\
&& {\cal L}_{{\bf g},\varkappa}^{ijk} = {\cal L}_{{\bf g},\varkappa}^{jik} \; , \qquad {\cal L}_{{\bf g},\varkappa}^{iik} = 0 \; ,\label{sym-tr0-k-g}\\
&& {\cal L}_{{\bf g},{\rm Ct}}^{ijk} = {\cal L}_{{\bf g},{\rm Ct}}^{jik} \; , \qquad
{\cal L}_{{\bf g},{\rm Ct}}^{iik} = 0 \; , \qquad
{\cal L}_{{\bf g},{\rm Cr}}^{ijk} = {\cal L}_{{\bf g},{\rm Cr}}^{jik}  \; , \qquad 
{\cal L}_{{\bf g},{\rm Cr}}^{iik} = 0 \; . \label{sym-tr0-C-g}
\eea

\vskip 0.2 cm

All these linear response coefficients can be explicitly obtained, as shown in the rest of this paper.

\subsection{The dissipative colloidal current densities}

Here, we focus on the translational and rotational dissipative current densities for dilute suspensions, where the colloidal chemical potential behaves as $\zeta_{\rm C}=\zeta_{\rm C}^0+k_{\rm B}T \ln (8\pi^2 f_{\rm C}/n_0)$, as shown in appendix~\ref{AppD}.  Since the density is much larger for the solvent than the colloidal species ($n_0 \gg n_{\rm C}$),  the inequality $\Vert\pmb{\nabla}f_{\rm C}/f_{\rm C}\Vert \gg \Vert\pmb{\nabla}n_0/n_0\Vert$ is satisfied.  Accordingly, the affinities driving the colloidal translational and rotational transports are given in dilute suspensions by
\be
\pmb{\mathscr A}_{\rm Ct} = - \frac{1}{f_{\rm C}} \, \pmb{\nabla} f_{\rm C} 
\qquad\mbox{and}\qquad
\pmb{\mathscr A}_{\rm Cr} = - \frac{1}{f_{\rm C}} \, \frac{\partial f_{\rm C}}{\partial\pmb{\alpha}} \, .
\label{A_Ct-A_Cr}
\ee

Now, we use the results of Brenner and coworkers \cite{B65,B67,BC72,HB83}, which are summarized in appendix~\ref{AppE}.  These results show that, for colloidal particles of arbitrary shape, the translational and rotational dissipative current densities take the following forms,
\bea
&&{\mathscr J}_{\rm Ct}^i = {\mathscr V}_{\rm t}^i \, f_{\rm C} - {\mathscr D}_{\rm t}^{ij} \, \nabla^j f_{\rm C} - {\mathscr D}_{\rm c}^{ji} \, \frac{\partial f_{\rm C}}{\partial \alpha^j} \, , \label{J_Ct-Brenner} \\
&&{\mathscr J}_{\rm Cr}^i = {\mathscr V}_{\rm r}^i \, f_{\rm C} - {\mathscr D}_{\rm c}^{ij} \, \nabla^j f_{\rm C} - {\mathscr D}_{\rm r}^{ij} \, \frac{\partial f_{\rm C}}{\partial \alpha^j} \, , \label{J_Cr-Brenner}
\eea
in terms of the translational and rotational dissipative drift velocities ${\mathscr V}_{\rm t}^i$ and ${\mathscr V}_{\rm r}^i$, the translational and rotational diffusion coefficients ${\mathscr D}_{\rm t}^{ij}={\mathscr D}_{\rm t}^{ji}$ and ${\mathscr D}_{\rm r}^{ij}={\mathscr D}_{\rm r}^{ji}$, and the coefficients ${\mathscr D}_{\rm c}^{ij}$ coupling translational and rotational colloidal diffusive transports.  The latter are non-zero for helicoidal particles \cite{B65,B67}.  The diffusion coefficients are known in the case of ellipsoidal particles, for which ${\mathscr D}_{\rm c}^{ij}=0$ \cite{P34,P36,B64b,B67,BC72}.  As a consequence, we obtain the following linear response coefficients,
\bea
&& {\mathscr L}_{{\rm Ct},{\rm Ct}}^{ij} = {\mathscr D}_{\rm t}^{ij} \, f_{\rm C} \, \delta(o-o') \; , \label{L_Ct,Ct-D_t}\\
&& {\mathscr L}_{{\rm Ct},{\rm Cr}}^{ij} = {\mathscr D}_{\rm c}^{ji} \, f_{\rm C} \, \delta(o-o') \; , \label{L_Ct,Cr-D_c}\\
&& {\mathscr L}_{{\rm Cr},{\rm Ct}}^{ij} = {\mathscr D}_{\rm c}^{ij} \, f_{\rm C} \, \delta(o-o') \; , \label{L_Cr,Ct-D_c}\\
&& {\mathscr L}_{{\rm Cr},{\rm Cr}}^{ij} = {\mathscr D}_{\rm r}^{ij} \, f_{\rm C} \, \delta(o-o') \; , \label{L_Cr,Cr-D_r}
\eea
in terms of the angular Dirac distribution
\be
\delta(o-o')\equiv \frac{1}{\sin\theta} \, \delta(\theta-\theta') \, \delta(\phi-\phi') \, \delta(\psi-\psi')
\label{Dirac_distrb}
\ee
and satisfying the Onsager reciprocal relations~(\ref{ORR_C}).

The dissipative drift velocities may have contributions from the reactions (r), the gradients of the molecular chemical potentials (d), and the gradients of the velocity field (g), as expressed according to
\bea
&&\pmb{\mathscr V}_{\rm t} = \pmb{\mathscr V}_{\rm t}^{\rm (r)} + \pmb{\mathscr V}_{\rm t}^{\rm (d)} + \pmb{\mathscr V}_{\rm t}^{\rm (g)} \, , \label{V_t}\\
&&\pmb{\mathscr V}_{\rm r} = \pmb{\mathscr V}_{\rm r}^{\rm (r)} + \pmb{\mathscr V}_{\rm r}^{\rm (d)} + \pmb{\mathscr V}_{\rm r}^{\rm (g)}= \pmb{\mathsf N}^{{\rm T}-1} \cdot \left( \pmb{\Omega}^{\rm (r)} + \pmb{\Omega}^{\rm (d)} + \pmb{\Omega}^{\rm (g)}\right) , \label{V_r}
\eea
in terms of the translational velocity $\pmb{\mathscr V}_{\rm t}^{\rm (x)}$ and the angular velocity $\pmb{\Omega}^{\rm (x)}$ of the contribution ${\rm x}\in\{{\rm r}, {\rm d}, {\rm g}\}$ given in Cartesian coordinates.

Accordingly, the contributions from the reactions are linearly related to the affinities of the chemical reactions by
\bea
&&{\mathscr V}_{\rm t}^{{\rm (r)}i} \, f_{\rm C}= \sum_r {\cal L}_{{\rm Ct},r}^i \, {\cal A}_{r} \, , \label{V_t^r}\\
&&{\mathscr V}_{\rm r}^{{\rm (r)}i} \, f_{\rm C}= \sum_r {\cal L}_{{\rm Cr},r}^i \, {\cal A}_{r} \, ; \label{V_r^r}
\eea
those from the molecular gradients to the corresponding affinities by
\bea
&&{\mathscr V}_{\rm t}^{{\rm (d)}i} \, f_{\rm C}= - \sum_\varkappa {\cal L}_{{\rm Ct},\varkappa}^{ij} \, \frac{\nabla^j n_\varkappa}{n_\varkappa} \, , \label{V_t^d}\\
&&{\mathscr V}_{\rm r}^{{\rm (d)}i} \, f_{\rm C}= - \sum_\varkappa {\cal L}_{{\rm Cr},\varkappa}^{ij} \, \frac{\nabla^j n_\varkappa}{n_\varkappa} \, ; \label{V_r^d}
\eea
and those from the velocity gradients to the corresponding affinities ${\mathsf A}_{\bf g}^{ij}=-{\mathsf S}^{ij}/(k_{\rm B}T)$, where ${\mathsf S}^{ij}\equiv\frac{1}{2}\left(\nabla^i v^j+\nabla^j v^i\right)$ is the symmetrised tensor of velocity gradients.  The latter contributions have been obtained in Refs.~\cite{B64b,B65,BC72}, showing that
\bea
&&{\mathscr V}_{\rm t}^{{\rm (g)}i} = - {\mathsf A}^{ijk} \, {\mathsf S}^{jk} \, , \label{V_t^g}\\
&&{\mathscr V}_{\rm r}^{{\rm (g)}i} = - \left(\pmb{\mathsf N}^{{\rm T}-1}\right)^{ij} \, {\mathsf B}^{jkl} \, {\mathsf S}^{kl} \, , \label{V_r^g}
\eea
in terms of the two triadics or rank-three tensors ${\mathsf A}^{ijk}$ and ${\mathsf B}^{ijk}$ (see appendix~\ref{AppE}).  Since ${\mathsf S}^{ij}={\mathsf S}^{ji}$, the triadics obey the symmetry relations ${\mathsf A}^{ijk}={\mathsf A}^{ikj}$ and ${\mathsf B}^{ijk}={\mathsf B}^{ikj}$.  Moreover, since the suspension is incompressible ${\mathsf S}^{ii}=0$, the triadics also satisfy ${\mathsf A}^{kii}=0$ and ${\mathsf B}^{kii}=0$.  As a consequence, we find that
\bea
&&{\cal L}_{{\rm Ct},{\bf g}}^{ijk} =  k_{\rm B}T \, {\mathsf A}^{ijk} \, f_{\rm C} \, , \label{L_Ct,g}\\
&&{\cal L}_{{\rm Cr},{\bf g}}^{ijk} =  k_{\rm B}T \, \left(\pmb{\mathsf N}^{{\rm T}-1}\right)^{il} \, {\mathsf B}^{ljk}  \, f_{\rm C} \, , \label{L_Cr,g}
\eea
which satisfy the Onsager-Casimir reciprocal relations~(\ref{CORR-C-g}), as well as the relations~(\ref{sym-tr0-C-g}).

\vskip 0.3 cm

We note that the diffusion coefficients can also be defined in suspensions and solutions that are not dilute, in which cases further couplings may arise \cite{GM62}.

\vskip 0.2 cm

An important remark is that the colloidal distribution function can be expanded as a series of Wigner functions of Euler's angles according to
\be
f_{\rm C}({\bf r},\pmb{\alpha},t) =\sum_{l,m',m} c_{m'm}^{(l)}({\bf r},t) \, D_{m'm}^{(l)}(\pmb{\alpha}) \, ,
\label{Wigner-series}
\ee
defining macrofields $c_{m'm}^{(l)}({\bf r},t)$ associated with specific angular distributions \cite{BP76,T11}.  These macrofields are ruled by partial differential equations that can be deduced from the local conservation equation~(\ref{eq-C-3}) for the colloidal particles by using the completeness and the orthogonality of the basis formed by the Wigner functions in the space of functions of Euler's angles.

\section{The case of spherical Janus particles}
\label{sec:Janus}

\subsection{Description of the suspension}

\begin{figure}[h]
\centerline{\scalebox{0.8}{\includegraphics{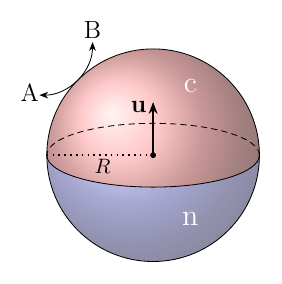}}}
\caption{Schematic representation of a Janus particle of radius $R$ with catalytic (c) and non-catalytic (n) hemispheres. The chemical reaction ${\rm A} \, \rightleftharpoons \, {\rm B}$ is catalysed by the c-hemisphere.
$\bf u$ denotes the unit vector along the axis of axisymmetry and pointing in the direction of the catalytic hemipshere.}
\label{fig2}
\end{figure}

Here, the colloidal particles are assumed to have a spherical shape of micrometric radius $R$ and a surface composed of two hemispheres: one being catalytic for the chemical reaction ${\rm A} \, \underset{-}{\stackrel{+}{\rightleftharpoons}} \, {\rm B}$ between two molecular species ${\rm A}$ and ${\rm B}$, which are present in the surrounding solution, and the other hemisphere being non-catalytic (i.e., inert for the reaction), as schematically depicted in figure~\ref{fig2}.  Such colloids are called Janus particles because of their two-faced structure \cite{CRRK14,RHSK16,GK19}.  They are axisymmetric and their orientation is given by the unit vector ${\bf u}=(\sin\theta\cos\phi,\sin\theta\sin\phi,\cos\theta)$ specified by the two angles $0\le\theta\le\pi$ and $0\le\phi <2\pi$ and pointing from the non-catalytic to the catalytic hemisphere.  Accordingly, the orientation of such axisymmetric particles does not need the third Eulerian angle $\psi$ to be specified.  Typically, the catalytic hemisphere is metallic and the non-catalytic one is made of polystyrene or silica.

In this suspension, there are thus the two reactive solute molecular species $\varkappa={\rm A}$ and $\varkappa={\rm B}$ in addition to the solvent $\varkappa=0$ and the colloidal particles ${\rm C}$.  There is no reaction occurring in the bulk of the solution.  The reaction ${\rm A} \, \underset{-}{\stackrel{+}{\rightleftharpoons}} \, {\rm B}$ occurs by heterogeneous catalysis at the interface between the solid material composing the catalytic hemisphere and the surrounding solution.  This reaction is characterised by the forward and backward rate constants $\kappa_{\pm}^{\rm c}$ defined per unit area and having the units $[\kappa_{\pm}^{\rm c}]=$~m/s.  In the suspension, the average rate constants of the reaction
\be
{\rm A} + {\rm C} \ \underset{-}{\stackrel{+}{\rightleftharpoons}} \ {\rm B} + {\rm C}
\label{reaction-AB}
\ee
are given by $k_{\pm}=4\pi R^2 a_0 \kappa_{\pm}^{\rm c}$ in terms of some dimensionless constant $a_0$ depending on the system \cite{GK18,GK19,GK20}.  In the absence of gradients of macrofields, the reaction rate density is entirely determined by the mass-action law according to
\be
w_r^{(0)} = w_{+r}^{(0)} - w_{-r}^{(0)} = k_+ \, n_{\rm A} \, n_{\rm C} - k_- \, n_{\rm B} \, n_{\rm C} =  - \sum_{\varkappa={\rm A},{\rm B}} \nu_\varkappa k_\varkappa \, n_\varkappa \, n_{\rm C} \, ,
\label{w_r-AB-0}
\ee
as expressed with the stoichiometric coefficients $\nu_{\rm A}=-1$ and $\nu_{\rm B}=+1$, and the rate constants $k_{\rm A}=k_+$ and $k_{\rm B}=k_-$.  This rate density has the units $[w_r]=$~m$^{-3}$~s$^{-1}$, as required.  The solution is dilute, so that the chemical afinity of the reaction is given by
\be
{\cal A}_r = \frac{\mu_{\rm A}-\mu_{\rm B}}{k_{\rm B}T} = \ln\frac{k_+ n_{\rm A}}{k_- n_{\rm B}} \, ,
\label{Aff-AB}
\ee
which is independent of the density $n_{\rm C}$ of the catalytic particles since $\nu_{\rm C}^{(+)}=\nu_{\rm C}^{(-)}$.
Close to chemical equilibrium, where $k_+ n_{\rm A,eq}=k_- n_{\rm B,eq}$, the affinity reads
\be
{\cal A}_r = \frac{1}{D_{\rm rxn}} \left(k_+ \delta n_{\rm A} - k_- \delta n_{\rm B}\right) ,
\qquad\mbox{where}\qquad
D_{\rm rxn} = k_+  n_{\rm A,eq} = k_- n_{\rm B,eq}
\label{Aff-AB-eq}
\ee
is the equilibrium diffusivity of the chemical reaction, having the units $[D_{\rm rxn}]=$~s$^{-1}$.  Therefore, the dependence of the reaction rate density~(\ref{w_r-AB-0}) on the chemical affinity has the following form
\be
w_r^{(0)} =  k_- \, n_{\rm B} \, n_{\rm C} \left({\rm e}^{{\cal A}_r} - 1 \right) = D_{\rm rxn} n_{\rm C} \, {\cal A}_r + O({\cal A}_r^2)
\label{w_r-AB-0-Aff}
\ee
and the linear response coefficient of the reaction is thus given by ${\cal L}_{r,r}=D_{\rm rxn} n_{\rm C}$.  We note that the gradients of the macrofields add contributions to the rate density~(\ref{w_r-AB-0}), as will be shown below.

The solute molecular species have the diffusion coefficients $D_{\rm A}$ and $D_{\rm B}$ of units $[D_\varkappa]=$~m$^2$/s, and the cross-diffusion between the molecular species is neglected. 

The suspension is assumed to be incompressible, $\pmb{\nabla}\cdot{\bf v}=0$, so that the dissipative part of the pressure tensor is traceless.  Moreover, it is symmetric by the local conservation of angular momentum.  The shear viscosity of the solution surrounding the colloids is denoted $\eta_0$.

Friction may occur at the interface between the solid particles and the solution.  This interfacial friction is characterised by a Navier slip length denoted $b$, which is supposed for simplicity to be uniform on the whole surface of the colloidal particles.

Moreover, diffusiophoresis is characterised by diffusiophoretic coefficients $b_\varkappa^h$ of units $[b_\varkappa^h]=$~m$^5$/s \cite{DD74,A89,GK18}.  They are specific of the molecular interaction forces between the solute species $\varkappa={\rm A},{\rm B}$ in the fluid and the material composing the hemispheric surface $h={\rm c},{\rm n}$ of the solid particle.  We may introduce their mean value and their difference measuring the diffusiophoretic heterogeneity between the two hemispheres as
\be
b_\varkappa \equiv \frac{1}{2} \left( b_\varkappa^{\rm c} + b_\varkappa^{\rm n}\right)
\qquad\mbox{and}\qquad
\Delta b_\varkappa \equiv b_\varkappa^{\rm c} - b_\varkappa^{\rm n} \, .
\label{b_k-Db_k}
\ee
If the diffusiophoretic coefficients were uniform on the whole surface of the Janus particles, their heterogeneity would be equal to zero: $\Delta b_\varkappa=0$.

The diffusiophoretic coefficients are further interfacial properties in addition to the surface rate constants $\kappa_{\pm}^{\rm c}$ (which are defined only on the catalytic hemisphere) and the Navier slip length $b$ (which is defined on the whole interface).  All these interfacial coefficients are assumed to take known values, which can in principle be determined from the microscopic atomic structure of the interface, using statistical mechanics \cite{RSGK20,G22,RSK24}.  These coefficients enter the chemohydrodynamic calculations at low Reynolds and P\'eclet numbers leading to the macroscopic current densities of the dilute active suspension on scales larger than the mean inter-particle distance, as shown in Refs.~\cite{GK18,GK19,GK20,G25}.  

\subsection{The local conservation equation for the spherical Janus particles}

For spherical colloidal particles, the translational and rotational diffusion processes are isotropic and there is no coupling between them, so that the rank-two tensors of colloidal diffusion are given by \cite{B64b,B67}
\be
{\mathscr D}_{\rm t}^{ij} = D_{\rm t} \, \delta^{ij} \, , \qquad
{\mathscr D}_{\rm r}^{ij} = D_{\rm r} \, g^{ij} \, , \qquad
{\mathscr D}_{\rm c}^{ij} = 0 \, , 
\label{diffusion-sph}
\ee
in terms of the following translational and rotational diffusion coefficients
\be
D_{\rm t} = \frac{k_{\rm B}T}{\gamma_{\rm t}}
\qquad\mbox{and}\qquad
D_{\rm r} = \frac{k_{\rm B}T}{\gamma_{\rm r}}
\label{Dt-Dr}
\ee
with the corresponding friction coefficients \cite{ABM75,F76b,LP08,GK18,G25}
\be
\gamma_{\rm t} = 6\pi\eta_0 R \, \frac{1+2b/R}{1+3b/R}
\qquad\mbox{and}\qquad
\gamma_{\rm r} = \frac{8\pi\eta_0 R^3}{1+3b/R} \, .
\label{frictions-t-r}
\ee

Since the particles are axisymmetric, their distribution function $f_{\rm C}({\bf r},\pmb{\alpha},t)$ is independent of the third Eulerian angle $\psi$, i.e., $\partial_\psi f_{\rm C}=0$.  As a consequence, the third component of the rotational vectors can be dropped and the configurational space of the colloids can be considered to be five-dimensional and given by the three-dimensional space~${\mathbb R}^3$ of the position $\bf r$ combined with the two-dimensional unit sphere ${\mathbb S}^2$ of the angular coordinates $\pmb{\alpha}=(\theta,\phi)$ needed for the unit vector ${\bf u}(\theta,\phi)$ \cite{GK20}.  In this case, the angular integration element should be redefined as $\ddo=\sin\theta\, \dd\theta \, \dd\phi$.  Consistently, the distribution function should also be redefined by replacing $\int f_{\rm C}\, \dd\psi=2\pi f_{\rm C}$ with $f_{\rm C}$, now having the units m$^{-3}$~rad$^{-2}$.

Accordingly, substituting equation~(\ref{diffusion-sph}) into equations~(\ref{J_Ct-Brenner}) and~(\ref{J_Cr-Brenner}), the local conservation equation~(\ref{eq-C-3}) with the translational and rotational current densities~(\ref{J_Ct-J_Cr-adv-diss}) is here given by
\be
\frac{\partial f_{\rm C}}{\partial t} + \pmb{\nabla}\cdot\Big(\pmb{\mathfrak V}_{\rm t} \, f_{\rm C} - D_{\rm t} \, \pmb{\nabla} f_{\rm C}\Big) + \frac{1}{\sin\theta} \, \frac{\partial}{\partial\theta}\left[ \sin\theta \left({\mathfrak V}_{\rm r}^{\theta} \, f_{\rm C} - D_{\rm r} \frac{\partial f_{\rm C}}{\partial\theta} \right) \right] + \frac{\partial}{\partial\phi}\left({\mathfrak V}_{\rm r}^{\phi} \, f_{\rm C} - \frac{D_{\rm r}}{\sin^2\theta} \, \frac{\partial f_{\rm C}}{\partial\phi} \right) = 0
\label{eq-C-sph}
\ee
in terms of the translational and rotational velocities
\be
\pmb{\mathfrak V}_{\rm t} = {\bf v} + \pmb{\mathscr V}_{\rm t} 
\qquad\mbox{and}\qquad
\pmb{\mathfrak V}_{\rm r} = \pmb{\upsilon}_{\rm r} + \pmb{\mathscr V}_{\rm r} \, ,
\label{velo-t-r}
\ee
including their reversible part due to advection and their dissipative part given by equations~(\ref{V_t})-(\ref{V_r}), with $\pmb{\mathfrak V}_{\rm t}=({\mathfrak V}_{\rm t}^x,{\mathfrak V}_{\rm t}^y,{\mathfrak V}_{\rm t}^z)$ and $\pmb{\mathfrak V}_{\rm r}=({\mathfrak V}_{\rm r}^\theta,{\mathfrak V}_{\rm r}^\phi)$.

As explained with equations~(\ref{V_t^r})-(\ref{V_r^g}), the colloidal particles can drift in the surrounding solution under the effects of several possible thermodynamic forces due to the reaction, the gradients of molecular concentrations, and the gradients of the velocity field.  However, for spherical particles, the rank-three tensors of coupling between colloidal transport and the velocity gradients are known to be equal to zero, ${\mathsf A}^{ijk}=0$ and ${\mathsf B}^{ijk}=0$ \cite{B64b}, so that the corresponding drift velocities (\ref{V_t^g})-(\ref{V_r^g}) and the corresponding coupling coefficients (\ref{L_Ct,g})-(\ref{L_Cr,g}) are also equal to zero.  Thus, there remain the contributions from the reaction and the gradients of molecular concentrations, which have been explicitly calculated in Refs.~\cite{GK18,GK19,GK20} for spherical Janus particles.  The result is that the translational and rotational dissipative drift velocities are given by
\bea
&&\pmb{\mathscr V}_{\rm t} = V_{\rm sd} \, {\bf u} + \sum_{\varkappa={\rm A},{\rm B}} \left( \xi_\varkappa \, \pmb{\mathsf I} + \varepsilon_\varkappa \, \pmb{\mathsf Q}_{\bf u} \right)\cdot \pmb{\nabla} n_\varkappa \, , \label{V_t-sph}\\
&&\pmb{\mathscr V}_{\rm r} = \sum_{\varkappa={\rm A},{\rm B}} \lambda_\varkappa \pmb{\nabla} n_\varkappa \cdot {\rm grad}_{\rm r} {\bf u}
\qquad\mbox{with}\qquad
{\rm grad}_{\rm r} {\bf u} = 
\left(
\begin{array}{c}
\frac{\partial{\bf u}}{\partial\theta} \\
\frac{1}{\sin^2\theta} \, \frac{\partial{\bf u}}{\partial\phi}
\end{array}
\right)
 \, , \label{V_r-sph}
\eea
in terms of the self-diffusiophoretic propulsion scalar velocity $V_{\rm sd}$, the parameters $\xi_\varkappa$, $\varepsilon_\varkappa$, and $\lambda_\varkappa$ of coupling to the molecular gradients, the $3\times 3$ unit tensor $\pmb{\mathsf I}=(\delta^{ij})$, the $3\times 3$ tensor of nematic order $\pmb{\mathsf Q}_{\bf u}=({\mathsf Q}_{\bf u}^{ij})=\left(u^i u^j-\frac{1}{3}\delta^{ij}\right)$, and the rotational gradient of the unit vector $\bf u$.

The propulsion velocity due to the reaction can be expressed as
\be
V_{\rm sd} = \chi \left( k_+ \, n_{\rm A} - k_- \, n_{\rm B} \right)
\label{V_sd}
\ee
in terms of the self-diffusiophoretic parameter
\be
\chi = \frac{a_1}{6\pi R^2 a_0 (1+2b/R)} \left[ \frac{b_{\rm B}}{D_{\rm B}} - \frac{b_{\rm A}}{D_{\rm A}} + O\left(\frac{\Delta b_\varkappa}{D_\varkappa} \right) \right] ,
\label{chi}
\ee
where $a_1$ is another dimensionless coefficients like $a_0$, which is specific of the system and which can be explicitly evaluated \cite{GK18,GK19,GK20}.  The coefficients $a_0$ and $a_1$ depend on the dimensionless Damk\"ohler number
\be
{\rm Da} \equiv R \left(\frac{\kappa_+^{\rm c}}{D_{\rm A}} + \frac{\kappa_-^{\rm c}}{D_{\rm B}}\right) ,
\label{Da}
\ee
characterising the reaction with respect to molecular diffusion.
The other parameters are also related to the diffusiophoretic coefficients or their heterogeneity \cite{GK20}:
\be
\xi_\varkappa = \frac{b_\varkappa}{1+2b/R} + O \left(\kappa_{\pm}^{\rm c} \frac{b_\varkappa^h}{D_\varkappa}R\right) , \qquad
\varepsilon_\varkappa = O \left(\kappa_{\pm}^{\rm c} \frac{b_\varkappa^h}{D_\varkappa}R\right) , \qquad
\mbox{and}\qquad
\lambda_\varkappa = O \left( \frac{\Delta b_\varkappa}{R}\right) + O \left(\kappa_{\pm}^{\rm c} \frac{\Delta b_\varkappa}{D_\varkappa}\right) .
\label{xi-eps-lambda}
\ee
The parameters $\xi_\varkappa$ directly characterise the diffusiophoresis of the colloidal particles in the molecular gradients.  The parameters $\varepsilon_\varkappa$ are furthermore proportional to the rate constants $\kappa_{\pm}^{\rm c}$ and the particle radius $R$ and, thus, they have a smaller effect.  The parameters $\lambda_\varkappa$ are due to heterogeneity in the diffusiophoretic coefficients $b_\varkappa^h$ between the two hemispheres and they would be equal to zero, if diffusiophoresis was homogeneous on the whole surface of the colloids (i.e., if $\Delta b_\varkappa=0$).  These parameters have the following units
\be
[\chi] = {\rm m} \, , \qquad [\xi_\varkappa] = {\rm m}^5/{\rm s} \, , \qquad [\varepsilon_\varkappa] = {\rm m}^5/{\rm s} \, , \qquad
\mbox{and}\qquad [\lambda_\varkappa] = {\rm m}^4/{\rm s} \, .
\label{units-parameters}
\ee

Close to chemical equilibrium, the self-diffusiophoretic scalar velocity (\ref{V_sd}) is related to the chemical affinity of the reaction by $V_{\rm sd}= \chi\, D_{\rm rxn} \, {\cal A}_r$. Accordingly, using equations~(\ref{V_t^r})-(\ref{V_r^g}), we obtain the following linear response coefficients:
\bea
&& {\cal L}_{{\rm Ct},r}^i = f_{\rm C} \, \chi \, D_{\rm rxn} \, u^i \, , \qquad {\cal L}_{{\rm Cr},r}^i = 0 \, , \label{L_Ct,r-L_Cr,r-sph}\\
&& {\cal L}_{{\rm Ct},\varkappa}^{ij} =  - n_\varkappa \,  f_{\rm C} \left(\xi_\varkappa \, \delta^{ij} + \varepsilon_\varkappa \, {\mathsf Q}_{\bf u}^{ij} \right) \, , \qquad {\cal L}_{{\rm Cr},\varkappa}^{ij} =  - n_\varkappa \,  f_{\rm C} \, \lambda_\varkappa \, {\rm grad}_{\rm r}^i u^j  \, , \label{L_Cr,k-L_Cr,k-sph}\\
&& {\mathscr L}_{{\rm Ct},{\rm Ct}}^{ij} = D_{\rm t} \, \delta^{ij} \, f_{\rm C} \, \delta(o-o') \, , \qquad
{\mathscr L}_{{\rm Cr},{\rm Cr}}^{ij} = D_{\rm r} \, g^{ij} \, f_{\rm C} \, \delta(o-o') \, , \qquad
{\mathscr L}_{{\rm Ct},{\rm Cr}}^{ij} = {\mathscr L}_{{\rm Cr},{\rm Ct}}^{ij} = 0 \, , \label{L_CC-sph} \\
&& {\cal L}_{{\rm Ct},{\bf g}}^{ijk} = 0 \, , \qquad {\cal L}_{{\rm Cr},{\bf g}}^{ijk} = 0 \, . \label{L_Cr,g-L_Cr,g-sph}
\eea

We note that nonlinear response coefficients could also be considered since the propulsion velocity~(\ref{V_sd}) calculated by chemohydrodynamics~\cite{GK18} can be written as $V_{\rm sd} = \chi \, k_- \, n_{\rm B} ({\rm e}^{{\cal A}_r}-1)$.

\subsection{The dissipative part of the pressure tensor}

Furthermore, the pressure tensor was calculated for active Janus particles of spherical shape in Ref.~\cite{G25}, showing that its dissipative part is given by
\be
\Pi^{ij} = - \eta\left(\nabla^i v^j + \nabla^j v^i \right) + \pi_r \, {\cal A}_r \, {\mathsf q}^{ij} - \sum_{\varkappa={\rm A},{\rm B}} \left[\tau_\varkappa \left( p^i \, \delta^{jk} + p^j \, \delta^{ik} -\frac{2}{3}\, p^k \, \delta^{ij}\right) + \varsigma_\varkappa \, {\mathsf r}^{ijk} \right] \nabla^k n_\varkappa \, ,
\label{Pi-sph}
\ee
where $p^i=\int u^i f_{\rm C}\, \ddo$, ${\mathsf q}^{ij}=\int {\mathsf Q}_{\bf u}^{ij} f_{\rm C}\, \ddo$, and ${\mathsf r}^{ijk}=\int {\mathsf R}_{\bf u}^{ijk} f_{\rm C}\, \ddo$ are the first, second, and third moments of the colloidal distribution function,
\be
\eta = \eta_0 \left( 1 + \frac{5}{2} \, \frac{1+2b/R}{1+5b/R} \, \Phi\right)
\qquad\mbox{with}\qquad
\Phi = \frac{4\pi}{3} R^3 n_{\rm C}
\label{eta-susp}
\ee
is the shear viscosity of the suspension including the correction due to the presence of the colloids \cite{E1906,LP08,PW22},
\be
\pi_r = -\frac{3\eta_0}{1+5b/R} \, \frac{a_2}{a_0} \left[ \frac{b_{\rm B}}{D_{\rm B}} - \frac{b_{\rm A}}{D_{\rm A}} + O\left(\frac{\Delta b_\varkappa}{D_\varkappa} \right) \right] D_{\rm rxn}
\label{pi_r}
\ee
is the parameter of coupling to the reaction, and
\bea
&& \tau_\varkappa = - \frac{3\eta_0}{1+5b/R} \, \frac{3(b_2+c_2)}{10 a_0} \left[ \frac{b_{\rm B}}{D_{\rm B}} - \frac{b_{\rm A}}{D_{\rm A}} + O\left(\frac{\Delta b_\varkappa}{D_\varkappa} \right) \right] \nu_\varkappa \, k_\varkappa R + O(\eta_0 \Delta b_\varkappa R^2) \, ,
\label{tau_k} \\
&& \varsigma_\varkappa = -\frac{3\eta_0}{1+5b/R} \frac{3b_2- 2 c_2}{2 a_0} \left[ \frac{b_{\rm B}}{D_{\rm B}} - \frac{b_{\rm A}}{D_{\rm A}} + O\left(\frac{\Delta b_\varkappa}{D_\varkappa} \right) \right] \nu_\varkappa \, k_\varkappa R + O(\eta_0 \Delta b_\varkappa R^2) \, ,
\label{sig_k}
\eea
are the parameters of coupling to the molecular concentration gradients.  The parameters~(\ref{pi_r}), (\ref{tau_k}), and~(\ref{sig_k}) are expressed in terms of dimensionless coefficients $a_2$, $b_2$, and $c_2$, which were evaluated in Ref.~\cite{G25} as functions of the Damk\"ohler number~(\ref{Da}).  These parameters have the following units,
\be
[\eta]=[\eta_0] = {\rm kg}\, {\rm m}^{-1} \, {\rm s}^{-1} \, , \qquad [\pi_r] = {\rm kg}\, {\rm m}^2 \, {\rm s}^{-2} \, , \qquad [\tau_\varkappa] = {\rm kg}\, {\rm m}^6 \, {\rm s}^{-2}\, , \qquad\mbox{and}\qquad [\varsigma_\varkappa] = {\rm kg}\, {\rm m}^6 \, {\rm s}^{-2} \, .
\ee

As a consequence of these results, we obtain the following linear response coefficients,
\bea
&& {\cal L}_{{\bf g},r}^{ij} = \pi_r \, q^{ij} \, , \label{L_g,r-sph}\\
&& {\cal L}_{{\bf g},\varkappa}^{ijk} = n_\varkappa \left[\tau_\varkappa \left( p^i \, \delta^{jk} + p^j \, \delta^{ik} -\frac{2}{3}\, p^k \, \delta^{ij}\right) + \varsigma_\varkappa \, {\mathsf r}^{ijk} \right] , \label{L_g,k-sph}\\
&& {\cal L}_{{\bf g},{\rm Ct}}^{ijk} = 0 \, , \qquad {\cal L}_{{\bf g},{\rm Cr}}^{ijk} = 0 \, , \label{L_g,Cr-L_g,Cr-sph}
\eea
and the elements ${\cal L}_{{\bf g},{\bf g}}^{ijkl}$ of the rank-four tensor of viscosity given in section~\ref{sec:dilute-suspension-A}, here with the shear viscosity~(\ref{eta-susp}).  The coefficients~(\ref{L_g,Cr-L_g,Cr-sph}) can also be deduced from the coefficients~(\ref{L_Cr,g-L_Cr,g-sph}) and the Onsager-Casimir reciprocal relations~(\ref{CORR-C-g}).

\subsection{The reaction rate density}

In addition to the leading contribution~(\ref{w_r-AB-0}) to the reaction rate density, there are further contributions caused by the gradients and its full expression is given by
\be
w_r = \left(k_+ \, n_{\rm A} - k_- \, n_{\rm B}\right) n_{\rm C} + \varpi \, p^i \left(k_+ \, \nabla^i n_{\rm A} - k_- \, \nabla^i n_{\rm B}\right) - \chi\, D_{\rm rxn} \, \nabla^i p^i + \frac{\pi_r}{k_{\rm B}T} \, {\mathsf q}^{ij} \, \nabla^i v^j
\label{w_r-AB}
\ee
in terms of the parameter $\varpi=O(R)$ with the units $[\varpi]=$~m, which was obtained in Ref.~\cite{GK20}, and the parameters~(\ref{chi}) and~(\ref{pi_r}).  We note that the term with the diffusiophoretic parameter $\chi$ comes from the coupling coefficient ${\cal L}_{{\rm Ct},r}^i$ in equation~(\ref{L_Ct,r-L_Cr,r-sph}) and the Onsager reciprocal relations ${\cal L}_{r,{\rm Ct}}^i={\cal L}_{{\rm Ct},r}^i$ in equation~(\ref{ORR_r}); and the term with the parameter $\pi_r$ from the coupling coefficient~(\ref{L_g,r-sph}) and the Onsager-Casimir reciprocal relations ${\cal L}_{r,{\bf g}}^{ij} = - {\cal L}_{{\bf g},r}^{ji}$ in equation~(\ref{CORR-r-g}).

As a consequence of equation~(\ref{w_r-AB}), we deduce the following linear response coefficients,
\be
{\cal L}_{r,\varkappa}^i = n_\varkappa \, \varpi \, p^i \, \nu_{\varkappa} \, k_\varkappa \, ,
\label{L_r,k-sph}
\ee
because $\nu_{\rm A}=-1$, $\nu_{\rm B}=+1$, $k_{\rm A}=k_+$, and $k_{\rm B}=k_-$.

\subsection{The dissipative part of the molecular current densities}

With all the previous results, the following full expression can be obtained for the dissipative current density of the molecular species $\varkappa$,
\be
{\cal J}_\varkappa^i = -D_\varkappa \, \nabla^i n_\varkappa + \varpi \, n_\varkappa \, \nu_{\varkappa} \, k_\varkappa \, {\cal A}_r \, p^i + n_\varkappa \left(\xi_\varkappa \, \nabla^i n_{\rm C} + \varepsilon_\varkappa \, \nabla^j {\mathsf q}^{ij}\right) + 2\, n_\varkappa \, \lambda_\varkappa \, p^i + \frac{n_\varkappa}{k_{\rm B}T} \left[\tau_\varkappa \, p^j \, (\nabla^i v^j + \nabla^j v^i) + \varsigma_\varkappa \, {\mathsf r}^{ijk} \, \nabla^j v^k \right] ,
\label{J_k-sph}
\ee
using
\be
\int \left(\xi_\varkappa\, \pmb{\mathsf I} + \varepsilon_\varkappa \, \pmb{\mathsf Q}_{\bf u}\right)\cdot\pmb{\nabla} f_{\rm C} \, \ddo= \xi_\varkappa\, \pmb{\nabla} n_{\rm C} + \varepsilon_\varkappa \, \pmb{\nabla}\cdot \pmb{\mathsf q}
\label{int-xi-eps-f}
\ee
and equation~(\ref{int-grad_u-grad_f}) of appendix~\ref{AppF}.  In equation~(\ref{J_k-sph}), the first term has the standard Fickian diffusion form, the second term is due to the coupling to the reaction by the reciprocals of the coefficients~(\ref{L_r,k-sph}), the third and fourth terms to the coupling to the gradients of the colloidal distribution function by the reciprocals of the coefficients~(\ref{L_Cr,k-L_Cr,k-sph}), and the fifth to the coupling to the velocity gradients by the reciprocals of the coefficients~(\ref{L_g,k-sph}) according to the Onsager-Casimir reciprocal relations~(\ref{CORR-k-g}).

\vskip 0.2 cm

Therefore, all the response coefficients of non-equilibrium thermodynamics are obtained by systematic calculations for active suspensions composed of spherical Janus particles.

\subsection{The entropy production rate density}

Gathering all the results, the entropy production rate density~(\ref{sigma_s}) for an isothermal and incompressible active suspension of spherical Janus particles is given by
\bea
\frac{1}{k_{\rm B}} \, \sigma_s &=& {\cal L}_{r,r} \, {\cal A}_r^2 +\sum_{\varkappa={\rm A},{\rm B}} D_\varkappa \, \frac{(\nabla^i n_\varkappa)^2}{n_\varkappa} + D_{\rm t} \int \frac{(\nabla^i f_{\rm C})^2}{f_{\rm C}}\, \ddo
+ D_{\rm r} \int \frac{g^{ij} }{f_{\rm C}}\, \frac{\partial f_{\rm C}}{\partial\alpha^i}\, \frac{\partial f_{\rm C}}{\partial\alpha^j}\, \ddo
+ \frac{\eta}{2 k_{\rm B}T}\, (\nabla^i v^j+\nabla^j v^i)^2 \nonumber\\
&& - \, 2 \, \varpi \, {\cal A}_r \sum_{\varkappa={\rm A},{\rm B}} \nu_\varkappa \, k_\varkappa \, \nabla^i n_\varkappa \int u^i \, f_{\rm C} \, \ddo - 2 \, \chi \, D_{\rm rxn} \, {\cal A}_r \int u^i \nabla^i f_{\rm C} \, \ddo
 \nonumber\\
&& - \, 2 \,  \sum_{\varkappa={\rm A},{\rm B}} \nabla^i n_\varkappa \int \Big( \xi_\varkappa \, \delta^{ij} + \varepsilon_\varkappa \, {\mathsf Q}_{\bf u}^{ij} \Big) \, \nabla^j f_{\rm C} \, \ddo - 2 \,  \sum_{\varkappa={\rm A},{\rm B}} \lambda_\varkappa \, \nabla^i n_\varkappa \int \frac{\partial u^i}{\partial \alpha^j} \, g^{jk} \, \frac{\partial f_{\rm C}}{\partial\alpha^k} \, \ddo \ge 0 \, ,
\label{sigma_s-sph}
\eea
where ${\cal L}_{r,r}=D_{\rm rxn} n_{\rm C}$.
The five leading terms are the contributions from the reaction, the diffusion of molecular species, the translational diffusion of colloids, the rotation diffusion of colloids, and viscosity.  The next terms arise from the couplings between these processes: the sixth term is due to the coupling between the reaction and molecular transport, the seventh term to the self-diffusiophoretic coupling between the reaction and the translational colloidal transport, the eighth to the coupling between molecular and translational colloidal transports, and the tenth to the coupling between molecular and rotational colloidal transports.  These coupling terms can generate energy transduction between the processes in the active suspension.  There is no contribution to the entropy production caused by the coupling to the velocity gradients because these couplings are odd under time reversal according to the Onsager-Casimir reciprocal relations~(\ref{CORR-r-g})-(\ref{CORR-C-g}).  For this reason, the coupling parameters $\pi_r$, $\tau_\varkappa$, and $\varsigma_\varkappa$ do not appear in the expression~(\ref{sigma_s-sph}) of the entropy production rate.    The non-negativity of the entropy production rate density~(\ref{sigma_s-sph}) requires the non-negativity of the coefficients of diffusivities and viscosity, as well as the following conditions among the transport coefficients and the coupling parameters of diffusiophoresis and self-diffusiophoresis:
\be
D_{\rm t} \gg \chi^2 \, D_{\rm rxn} \, , \quad
D_{\rm t} \, D_\varkappa \gg n_{\rm C} \, n_\varkappa \, \xi_\varkappa^2 \, , \quad
D_{\rm t} \, D_\varkappa \gg n_{\rm C} \, n_\varkappa \, \varepsilon_\varkappa^2 \, , \quad
D_{\rm r} \, D_\varkappa \gg n_{\rm C} \, n_\varkappa \, \lambda_\varkappa^2 \, , \quad
D_{\rm rxn} \, D_\varkappa  \gg \varpi^2 \, n_{\rm C} \, n_\varkappa \, k_\varkappa^2 \, ,
\label{inequalities}
\ee
as shown in appendix~\ref{AppH}.

We note that, if the chemical reaction runs in the nonlinear regime, its contribution ${\cal L}_{r,r} \, {\cal A}_r^2$ to the entropy production rate density~(\ref{sigma_s-sph}) should be replaced by ${\cal L}_{r,r} \, {\cal A}_r\left({\rm e}^{{\cal A}_{\rm r}}-1\right)$ according to equation~(\ref{w_r-AB-0-Aff}).  Other contributions could be extended similarly.

\subsection{Dynamics of the moments of the colloidal distribution function}

Now, we can derive a hierarchy of evolution equations for the successive moments of the colloidal distribution function
from its local conservation equation~(\ref{eq-C-sph}).  These moments are systematically defined in appendix~\ref{AppF}.  Accordingly, the colloidal distribution function can be expanded as
\be
f_{\rm C}({\bf r},\pmb{\alpha},t) = \frac{1}{4\pi} \left[ n_{\rm C}({\bf r},t) + 3\, {\bf u}\cdot{\bf p}({\bf r},t) + \frac{15}{2}\, \pmb{\mathsf Q}_{\bf u} : \pmb{\mathsf q}({\bf r},t) + \cdots \right] ,
\label{expansion-f_C-truncated}
\ee
where $n_{\rm C}=\int f_{\rm C}\, \ddo$ is the colloidal density, ${\bf p}=\int {\bf u} \, f_{\rm C}\, \ddo$ is the vector of average polar order, $\pmb{\mathsf q}=\int \pmb{\mathsf Q}_{\bf u} \, f_{\rm C}\, \ddo$ is the rank-two tensor of average nematic order, and the dots denote higher-order terms.

The zeroth-order moment, i.e., the colloidal density of Janus particles $n_{\rm C}$, is ruled by the local conservation equation~(\ref{eq-n_C}) here given by
\be
\partial_t \, n_{\rm C} + \pmb{\nabla}\cdot \pmb{\jmath}_{\rm C} = 0
\qquad\mbox{with}\qquad
\pmb{\jmath}_{\rm C} = n_{\rm C} \, {\bf v} + V_{\rm sd} \, {\bf p} + n_{\rm C} \sum_{\varkappa={\rm A},{\rm B}} \xi_\varkappa \, \pmb{\nabla} n_\varkappa + \sum_{\varkappa={\rm A},{\rm B}} \varepsilon_\varkappa \, \pmb{\mathsf q}\cdot \pmb{\nabla} n_\varkappa - D_{\rm t} \, \pmb{\nabla} n_{\rm C} \, .
\label{eq-n_C-sph}
\ee

Next, the evolution equation for the vector of average polar order can be obtained from the local conservation equation~(\ref{eq-C-sph}).  The calculation is carried out in appendix~\ref{AppG}.  The result is given by equations~(\ref{DpDt}) and~(\ref{eq-p}).  In the case where there is no diffusiophoretic heterogeneity (i.e., $\Delta b_\varkappa=0$ so that $\lambda_\varkappa=0$), the effects of $O(R)$ are negligible (i.e., $\varpi=0$ and $\varepsilon_\varkappa=0$), and the higher moments are absent (i.e., ${\mathsf q}^{ik}=0$, ${\mathsf r}^{ijk}=0$,$\dots$), the vector of polar order should evolve according to
\be
\partial_t p^i + v^j\nabla^j p^i + \frac{1}{2}(\nabla^i v^j -\nabla^j v^i) p^j = -\frac{1}{3} \nabla^i(V_{\rm sd} \, n_{\rm C}) -\nabla^j\bigg( p^i\nabla^j\sum_{\varkappa={\rm A},{\rm B}}\xi_\varkappa n_\varkappa \bigg) + D_{\rm t} \nabla^2 p^i - 2 D_{\rm r} p^i \, .
\label{eq-p-approx}
\ee
The last term expresses the relaxation of polar order at the rate $2D_r$ due to rotational diffusion.
If this diffusion was fast enough, the polar-order vector would be approximately given by ${\bf p}\simeq (6D_{\rm r})^{-1}\pmb{\nabla}(V_{\rm sd}\, n_{\rm C})$.  Under such circumstances, using equation~(\ref{int-nabla_f-nabla_f}) and~(\ref{int-grad_f-grad_f}), the entropy production rate density could be evaluated according to
\bea
\frac{1}{k_{\rm B}} \, \sigma_s &=& {\cal L}_{r,r} \, {\cal A}_r^2 +\sum_{\varkappa={\rm A},{\rm B}} D_\varkappa \, \frac{(\pmb{\nabla} n_\varkappa)^2}{n_\varkappa} + \frac{\eta}{2 k_{\rm B}T}\, (\pmb{\nabla}{\bf v}+\pmb{\nabla}{\bf v}^{\rm T})^2 \nonumber\\
&& + \frac{D_{\rm t}}{n_{\rm C}} \left[ (\pmb{\nabla} n_{\rm C})^2 + 3 \left(\nabla^i p^j - p^j \frac{\nabla^i n_{\rm C}}{n_{\rm C}}\right)^2 \right] + \frac{6 D_{\rm r}}{n_{\rm C}}\, {\bf p}^2 \nonumber\\
&& - 2 \, \chi \, D_{\rm rxn} \, {\cal A}_r \, \pmb{\nabla}\cdot{\bf p} 
- 2  \sum_{\varkappa={\rm A},{\rm B}} \xi_\varkappa \pmb{\nabla} n_\varkappa \cdot \pmb{\nabla} n_{\rm C} \, ,
\label{sigma_s-sph-approx}
\eea
up to higher-order terms.

The evolution equation for the tensor $\pmb{\mathsf q}$ of nematic order can also be obtained, as shown in appendix~\ref{AppG}, as well as its contributions to the entropy production rate density~(\ref{sigma_s-sph}).

\section{Conclusion and perspectives}
\label{sec:conclusion}

In this paper, the non-equilibrium thermodynamics of active suspensions was developed starting from the extension of thermodynamics to systems with internal degrees of freedom by Prigogine and Mazur~\cite{PM53}.  In active suspensions, these degrees of freedom are the Eulerian angles specifying the orientation of the colloidal particles.  Accordingly, their local conservation equation should be formulated in the six-dimensional configuration space of their position and their orientation.  Moreover, the chemical reactions powering the activity of the suspension were also taken into account.  These reactions occur by heterogeneous catalysis at the surface of the colloidal particles, which are thus propelled through the surrounding fluid by self-diffusiophoresis, i.e., the mechanism coupling the molecular concentration gradients to the velocity field of the suspension.  

On this basis, the local balance equation for entropy was derived from the local conservation laws for the colloids, mass, linear momentum, and energy, combined with the rules of chemical kinetics for the molecular species.  In this way, the list of processes contributing to the entropy production rate of active suspensions was obtained.  This list includes the chemical reactions, the translational and rotational transports of colloidal particles, in addition to the standard transport processes of molecular species, heat, and linear momentum in shear and dilatational flows.  The contribution of each process is given by the product of the associated thermodynamic force or affinity multiplied with the corresponding dissipative current density, which are each clearly identified by the theory (see table~\ref{Tab1}).

Next, the non-equilibrium constitutive relations were established between the dissipative current densities and the affinities in isothermal, incompressible, dilute suspensions, including all the possible couplings between the processes in consistency with the Onsager-Casimir reciprocal relations implied by microreversibility.  The extension to the internal degrees of freedom of the catalytic colloidal particles here plays an essential role to obtain the mechanochemical couplings powering the non-equilibrium activity of the system.  The compatibility with Curie's symmetry principle is satisfied because the catalytic colloids have internal degrees of freedom corresponding to their orientation.  Accordingly, the extension of Curie's symmetry principle to the six-dimensional configuration space of the colloidal particles allows new possible couplings, which are otherwise forbidden in usual isotropic reactive multicomponent solutions.  Furthermore, the knowledge of the colloidal distribution function and its dynamics completely determines the time evolution of the polar, nematic, and higher orientational orders, which can be directly deduced within the framework of the present theory without needing any further empirical assumptions.

Remarkably, all the coupling coefficients can be calculated in the case of spherical Janus particles formed by a catalytic hemisphere and another intert one, as carried out in previous work \cite{GK18,GK19,GK20,G25}.  Because of their micrometric size, the colloidal particles are much larger than the atoms and molecules composing the system.  Therefore, the reactions and the movements of the colloids can be described using fluid mechanics coupled to the advection-diffusion-reaction equations for the molecular species in terms of the geometry of the particles and the interfacial properties coupling the fluid velocity to the molecular concentration fields in the solution around the particles.  The interfacial properties are the surface reaction rate constants of the catalytic hemisphere, the diffusiophoretic coefficients related to the molecular interaction forces between the solid materials composing the particles and the molecules in the surrounding solution, and the Navier slip length characterising the friction between the fluid and the solid surface.  In the limit of a dilute suspension, the colloidal particles participate to the chemohydrodynamics of the active suspension mainly through the mediation of the fluid solution, in which the particles are immersed, since the direct mechanical interactions between the particles are rare in this limit.  Consequently, the macroscopic non-equilibrium constitutive relations of such dilute active suspensions can be completely calculated and predicted by the theoretical methods of chemohydrodynamics, after averaging over the statistical distribution of colloidal particles at the macroscale.  In particular, the analytic expression is obtained for the entropy production rate of such an active suspension.  

We note that, although all the couplings are possible because of microreversibility, they do not all have contributions of equal magnitudes in the dynamics of the active suspension.  Indeed, some of the contributions to a particular dissipative current density may be negligible under non-equilibrium conditions, so that some of the reciprocal effects predicted by microreversibility could be effectively ignored \cite{GK20}.  Such circumstances may lead us to consider non-reciprocal approximations for the description of active matter.

Away from equilibrium, the chemical reactions can give important contributions to the entropy production rate, especially, since they drive the activity of the suspension.  Therefore, they play crucial roles in the mechanisms of energy interconversion powering active matter.  These mechanisms are due to the couplings established by the methods of non-equilibrium thermodynamics.  In this regard, the knowledge of the entropy production rate allows us to evaluate the thermodynamic costs and efficiencies of these mechanisms.

All these results open important perspectives in our understanding of active suspensions and, more generally, active matter.  Let us now mention several issues left open in the present work.

A first open issue is to formulate non-equilibrium thermodynamics for active suspensions subjected to an external force field such a gravitational or an electrical external field.  In such systems, external forces and torques are exerted on the colloidal particles, so that the local conservation equation for linear momentum should be extended to include a source term with the external force density and the pressure tensor should have a non-symmetric contribution due to external torque.

Another issue is to calculate the properties of active suspensions composed of colloids with other shapes than with spherical Janus particles.  The particles may be ellipsoidal or helicoidal, which may induce new types of coupling between the transport processes \cite{B64b,B65,B67,BC72,PDTR10}.  Moreover, the active suspension may be polydisperse with colloids of different sizes or different shapes, or may be a mixture of several kinds of active or passive colloidal particles \cite{CT18}.  The active particles or agents may also have moving parts contributing to their propulsion and the present theoretical methods give hints to understand how their internal degrees of freedom can be taken into account.  There also exist other phoretic effects like electrodiffusiophoresis or thermophoresis.  The latter effect can generate activity in systems heated by an electromagnetic radiation field \cite{A89,BDLRVV16}.  The dispersion medium may be a gas rather than a liquid, in which case the particles are aerosols.  If the dispersion medium is a plasma, the system is a dusty plasma, existing in planetary nebulae and possibly the stage of reactions catalysed by the dust particles \cite{MI09,Plasmas2023}.

An important issue is to extend the present results to denser active suspensions.  For this purpose, there exist systematic theoretical methods to calculate properties such as the effective viscosity of a suspension as power series of the colloidal density \cite{BKM77,HB83}.  These methods can also be used to expand the properties in powers of the response frequency to go beyond the quasistatic approximation.  We can also envisage expansions in powers of the Reynolds number beyond the laminar regime of hydrodynamics or in powers of the P\'eclet numbers to include corrections due to the advection of molecular species, which may lead to intertwined effects between the hydrodynamic and chemotactic interactions between the colloidal particles.  Furthermore, in dense enough suspensions, there exist transitions towards phase separations or colloidal crystalline phases, which can be described with suitable non-linear constitutive relations beyond the linear constitutive relations considered for dilute suspensions, in particular, using Ginzburg-Landau or Cahn-Hilliard free energy functionals \cite{GL50,CH58}.  Motility-induced phase separation may also happen in dense systems of self-propelled particles \cite{CT15}.  Moreover, the micrometric particles can also evolve under the effect of the chemical reactions.

The present theoretical considerations about internal degrees of freedom open new perspectives in the study of other sorts of active matter with different forms of mesostructure or mesotexture.  Future work will address the issue of the chemohydrodynamic instabilities that are the consequences of the coupling mechanisms predicted by the non-equilibrium thermodynamics of active suspensions.  Such instabilities have already been investigated if the solution is at rest with a zero fluid velocity.  Otherwise, convective instabilities may also be induced in active suspensions.

More generally speaking, we note that the results of non-equilibrium thermodynamics can be used as guidelines for microscopic approaches based on non-equilibrium statistical mechanics.  In conclusion, non-equilibrium thermodynamics constitutes a powerful and predictive theoretical framework, which is highly important for the study of active matter.

\section*{Acknowledgments}

This research was supported by the Universit\'e Libre de Bruxelles (ULB).
The author thanks Raymond Kapral for discussions, Jo\"el Mabillard for his careful reading of the draft and his many helpful comments, and David Gaspard for his expertise.

\vskip 0.5 cm

\appendix

\section{Rigid-body orientation and torque}
\label{AppA}

\subsection{The Eulerian angles}

The orientation of a rigid body of arbitrary shape can be specified with three Eulerian angles $\pmb{\alpha}=(\theta,\phi,\psi)$ corresponding to three successive rotations around given axes, bringing the reference to the current orientation.  Here, we choose the three successive rotations in order for the unit vector ${\bf 1}_z=(0,0,1)$ to be mapped onto the unit vector ${\bf u}=(\sin\theta\cos\phi,\sin\theta\sin\phi,\cos\theta)$.  This is achieved using the $3\times 3$ orthogonal matrix given by
\be
\pmb{\mathsf O}(\pmb{\alpha}) \equiv \pmb{\mathsf O}_z(\phi)\cdot\pmb{\mathsf O}_y(\theta)\cdot\pmb{\mathsf O}_z(\psi) \, ,
\label{rot-O}
\ee
where
\bea
&& \pmb{\mathsf O}_z(\psi) = 
\left(
\begin{array}{ccc}
\cos\psi & -\sin\psi & 0 \\
\sin\psi & \cos\psi & 0 \\
0 & 0 & 1
\end{array}
\right)
\qquad\mbox{with}\qquad 0 \le \psi < 2\pi \, , 
\label{rot-O-z-psi}
\\
&& \pmb{\mathsf O}_y(\theta) = 
\left(
\begin{array}{ccc}
\cos\theta & 0 & \sin\theta \\
0 & 1 & 0 \\
-\sin\theta & 0 & \cos\theta
\end{array}
\right)
\qquad\mbox{with}\qquad 0 \le \theta \le \pi \, , 
\label{rot-O-y-theta}
\\
&& \pmb{\mathsf O}_z(\phi) = 
\left(
\begin{array}{ccc}
\cos\phi & -\sin\phi & 0 \\
\sin\phi & \cos\phi & 0 \\
0 & 0 & 1
\end{array}
\right)
\qquad\mbox{with}\qquad 0 \le \phi < 2\pi \, .
\label{rot-O-z-phi}
\eea

The rotation~(\ref{rot-O}) maps any vector ${\bf R}_0$ of the initial reference orientation onto the vector ${\bf R}=\pmb{\mathsf O}(\pmb{\alpha})\cdot{\bf R}_0$ of the orientation at current time, so that its differential can be expressed as
\be
\dd{\bf R}=\dd\pmb{\mathsf O}(\pmb{\alpha})\cdot{\bf R}_0= \dd\pmb{\alpha}\cdot \partial_{\pmb{\alpha}}\pmb{\mathsf O}(\pmb{\alpha})\cdot \pmb{\mathsf O}^{-1}(\pmb{\alpha})\cdot{\bf R} \equiv \dd\pmb{\chi}\times{\bf R} \, ,
\label{dR-dchi}
\ee
which implies that
\be
\dd\pmb{\chi} = \pmb{\mathsf N}^{\rm T}\cdot\dd\pmb{\alpha} 
\qquad\mbox{with}\qquad
{\mathsf N}^{ij} =\frac{1}{2}\, \epsilon^{jkl} \, {\mathsf O}^{km} \, \frac{\partial{\mathsf O}^{lm}}{\partial\alpha^i} \, .
\label{dfn-matrix-N}
\ee
The $3\times 3$ matrix $\pmb{\mathsf N}$ plays a central role in the mechanics of the rigid body~\cite{RSGK20} and, also, in the theory of rotational Brownian motion \cite{B67}.  Here, it is explicitly given by
\be
\pmb{\mathsf N} = 
\left(
\begin{array}{ccc}
-\sin\phi & \cos\phi & 0 \\
0 & 0 & 1 \\
\sin\theta\cos\phi & \sin\theta\sin\phi & \cos\theta
\end{array}
\right) .
\label{matrix-N}
\ee
This matrix is the Jacobian matrix of the transformation of the vector $\dd\pmb{\alpha}$ into $\dd\pmb{\chi}$.
Its determinant $\det\pmb{\mathsf N}=\sin\theta$ is the Jacobian determinant giving the element of integration over the Eulerian angles: $\dd o = \dd^3\chi = (\det\pmb{\mathsf N}) \dd^3\alpha = \sin\theta\, \dd\theta\, \dd\phi \, \dd\psi$.  Using Euler's angles $\pmb{\alpha}$ as curvilinear coordinates in the vectorial space spanned by the vectors $\dd\pmb{\chi}$, the metric is given by $\dd\pmb{\chi}^2 = \pmb{\mathsf g}:\dd\pmb{\alpha}^2=g_{ij}\, \dd\alpha^i \, \dd\alpha^j$ with
\be
(g_{ij}) = \pmb{\mathsf g} = \pmb{\mathsf N}\cdot\pmb{\mathsf N} ^{\rm T} =
\left(
\begin{array}{ccc}
1 & 0 & 0 \\
0 & 1 & \cos\theta \\
0 & \cos\theta & 1
\end{array}
\right) ,
\label{matrix-g}
\ee
so that $\dd\pmb{\chi}^2 =\dd\theta^2 + \dd\phi^2 + \dd\psi^2 + 2 \cos\theta\, \dd\phi \, \dd\psi$.  The integration element can be written as $\dd o = \sqrt{g}\, \dd^3\alpha$ in terms of the determinant of the metric: $g\equiv\det\pmb{\mathsf g}=(\det\pmb{\mathsf N})^2=\sin^2\theta$.
The inverse of the metric~(\ref{matrix-g}) reads
\be
(g^{ij}) = \pmb{\mathsf g}^{-1} = \pmb{\mathsf N}^{{\rm T}-1}\cdot\pmb{\mathsf N} ^{-1} =
\left(
\begin{array}{ccc}
1 & 0 & 0 \\
0 & \frac{1}{\sin^2\theta} & -\frac{\cos\theta}{\sin^2\theta} \\
0 & -\frac{\cos\theta}{\sin^2\theta} & \frac{1}{\sin^2\theta}
\end{array}
\right) .
\label{matrix-inv-g}
\ee

\subsection{Torque and potential energy in Eulerian angles}

If the rigid body is subjected to a potential energy $U$ depending on its orientation, the torque exerted on it is given by
\be
{\bf T} = {\bf R}\times {\bf F} 
\qquad\mbox{with}\qquad
{\bf F} = - \frac{\partial U}{\partial{\bf R}} \, ,
\label{toruqe}
\ee
which can be expressed in terms of the Eulerian angles as follows.  Since the vector $\bf R$ is related to the Eulerian angles, we have that
\be
T^i = \epsilon^{ijk} \, R^j \, F^k = - \epsilon^{ijk} \, R^j \, \frac{\partial U}{\partial R^k} = - \epsilon^{ijk} \, R^j \, \frac{\partial U}{\partial\alpha^l} \, \frac{\partial \alpha^l}{\partial R^k}
\label{torque-Cart}
\ee
in Cartesian coordinates.
Now, because of equations~(\ref{dR-dchi}) and~(\ref{dfn-matrix-N}), the infinitesimal increments $\dd{\bf R}$ and $\dd\pmb{\alpha}$ are related by $\dd R^i =\epsilon^{ijk} R^k {\mathsf N}^{lj}\dd\alpha^l$, so that the Jacobian matrix of the transformation from the coordinates $\bf R$ to the Eulerian angles $\pmb{\alpha}$ is given by
\be
\frac{\partial R^j}{\partial\alpha^i} = \epsilon^{jkl} R^l \, {\mathsf N}^{ik} \, .
\label{dR-dalpha}
\ee
Since
\be
\frac{\partial U}{\partial\alpha^i} = \frac{\partial U}{\partial R^j} \, \frac{\partial R^j}{\partial\alpha^i} = - F^j \, \frac{\partial R^j}{\partial\alpha^i}
\ee
we find that
\be
\frac{\partial U}{\partial\pmb{\alpha}} = - \pmb{\mathsf N} \cdot{\bf T} 
\qquad\mbox{and, thus,}\qquad
{\bf T} = - \pmb{\mathsf N}^{-1} \cdot \frac{\partial U}{\partial\pmb{\alpha}} \, .
\label{dU-torque}
\ee

\section{Vector calculus with Eulerian angles}
\label{AppB}

The relation~(\ref{J_Cr}) can be justified as follows.  For any function $f$ depending on the orientation, we should have that
\be
\pmb{\mathfrak J}_{\rm Cr}^{\rm T} \cdot\frac{\partial f}{\partial\pmb{\alpha}} = {\bf J}_{\rm Cr}^{\rm T} \cdot \frac{\partial f}{\partial\pmb{\chi}} 
\qquad\mbox{or, equivalently,}\qquad
{\mathfrak J}_{\rm Cr}^i \, \frac{\partial f}{\partial\alpha^i} = {\rm J}_{\rm Cr}^j \, \frac{\partial f}{\partial\chi^j} \, .
\ee
Because of equation~(\ref{dfn-matrix-N}), we get
\be
{\mathfrak J}_{\rm Cr}^i \, \frac{\partial \chi^j}{\partial\alpha^i} \, \frac{\partial f}{\partial\chi^j} ={\mathfrak J}_{\rm Cr}^i \, {\mathsf N}^{ij} \, \frac{\partial f}{\partial\chi^j} \, ,
\ee
so that
\be
\pmb{\mathsf N}^{\rm T} \cdot \pmb{\mathfrak J}_{\rm Cr} = {\bf J}_{\rm Cr} \, ,
\label{J_Cr-T-N}
\ee
leading to equation~(\ref{J_Cr}) after inverting $\pmb{\mathsf N}^{\rm T}$.

\vskip 0.2 cm

The rotational gradient of some function $f$ is defined by
\be
{\rm grad}_{\rm r} f \equiv \pmb{\mathsf g}^{-1}\cdot\frac{\partial f}{\partial\pmb{\alpha}}
\qquad\mbox{or, equivalently,}\qquad
{\rm grad}_{\rm r}^i f \equiv g^{ij}\, \frac{\partial f}{\partial\alpha^j}
\label{dfn-grad_r}
\ee
in terms of the inverse of the metric expressed by equation~(\ref{matrix-inv-g}).  Therefore, the components of the rotational gradient are given by
\be
{\rm grad}_{\rm r} f = 
\left[
\begin{array}{c}
\partial_\theta f \\
\frac{1}{\sin^2\theta} \left(\partial_\phi f - \cos\theta \, \partial_\psi f \right) \\
\frac{1}{\sin^2\theta} \left(-\cos\theta \, \partial_\phi f  + \partial_\psi f \right)
\end{array}
\right] .
\label{components-grad_r}
\ee

\vskip 0.2 cm

Taking the rotational divergence~(\ref{dfn-div-r}) of the rotational gradient, we obtain the rotational Laplacian in Eulerian angles as
\be
\hat{\cal L}_{\rm r}f \equiv {\rm div}_{\rm r} \, {\rm grad}_{\rm r} f = \frac{1}{\sin\theta} \, \partial_\theta\left(\sin\theta \, \partial_\theta f \right) + \frac{1}{\sin^2\theta} \left( \partial_\phi^2 f + \partial_\psi^2 f - 2 \cos\theta \, \partial_\phi\partial_\psi f \right) ,
\label{rot-Laplacian}
\ee
which describes isotropic rotational diffusion.

\vskip 0.2 cm

For axisymmetric particles having an orientation specified only by the two angles $\pmb{\alpha}=(\theta,\phi)$, the function $f$ is independent of the third Eulerian angle $\psi$, i.e., $\partial_\psi f=0$.  As a consequence, the third term of the rotational divergence~(\ref{dfn-div-r}) is equal to zero.  Therefore, the third component of the rotational gradient no longer plays any role and can be dropped.  In this case, we can use the following rotational gradient with only two components,
\be
{\rm grad}_{\rm r} f = 
\left(
\begin{array}{c}
\partial_\theta f \\
\frac{1}{\sin^2\theta} \, \partial_\phi f
\end{array}
\right) ,
\qquad\mbox{if} \qquad \partial_\psi f= 0 \, ,
\label{2_components-grad_r}
\ee
which can be expressed as in equation~(\ref{dfn-grad_r}) but with the $2\times 2$ matrix 
\be
(g_{ij}) = \pmb{\mathsf g} =
\left(
\begin{array}{cc}
1 & 0  \\
0 & \sin^2\theta
\end{array}
\right) ,
\qquad\mbox{if} \qquad \partial_\psi f= 0 \, ,
\label{2-matrix-g}
\ee
such that $\dd{\bf u}^2 = g_{ij}\, \dd\alpha^i \, \dd\alpha^j=\dd\theta^2 + \sin^2\theta \, \dd\phi^2$.  Accordingly, the rotational Laplacian~(\ref{rot-Laplacian}) reduces to
\be
\hat{\cal L}_{\rm r}f \equiv {\rm div}_{\rm r} \, {\rm grad}_{\rm r} f = \frac{1}{\sin\theta} \, \partial_\theta\left(\sin\theta \, \partial_\theta f \right) + \frac{1}{\sin^2\theta} \, \partial_\phi^2 f \, ,
\qquad\mbox{if}\qquad
\partial_\psi f = 0 \, .
\label{rot-Laplacian-2}
\ee

\section{Derivation of the local balance equation for entropy}
\label{AppC}

The time derivative along stream lines in the space of position and orientation is given by
\be
\frac{\dd}{\dd t} = \frac{\partial}{\partial t} + {\bf v}\cdot\pmb{\nabla} + \pmb{\upsilon}_{\rm r}\cdot\frac{\partial}{\partial\pmb{\alpha}} \, ,
\label{d-dt}
\ee
where $\bf v$ is the velocity field, i.e., the velocity in position space, and $\pmb{\upsilon}_{\rm r}$ is the velocity~(\ref{v_r}) in the orientation space.  This time derivative is supposed to apply to macrofields.  We note that the last term is equal to zero if the macrofield is independent of the Eulerian angles $\pmb{\alpha}$.

Along stream lines, the local Gibbs thermodynamic relation~(\ref{Gibbs}) reads
\be
\frac{\dd s}{\dd t} = \frac{1}{T} \, \frac{\dd e}{\dd t} - \sum_{\varkappa = 0}^{\mathscr M} \frac{\mu_\varkappa}{T} \, \frac{\dd n_\varkappa}{\dd t} - \int \ddo \, \frac{\zeta_{\rm C}}{T} \, \frac{\dd f_{\rm C}}{\dd t} \, .
\label{Gibbs-ddt}
\ee

Using the local conservation equation~(\ref{eq-mass}) for mass, equation~(\ref{eq-g}) for linear momentum becomes
\be
\rho\left(\partial_t+{\bf v}\cdot\pmb{\nabla}\right){\bf v} = -\pmb{\nabla}\cdot\pmb{\mathsf P} \, ,
\label{eq-g-2}
\ee
where the pressure tensor is symmetric $\pmb{\mathsf P}=\pmb{\mathsf P}^{\rm T}$.
Next, using equations~(\ref{eq-mass}) and~(\ref{eq-g-2}), the local conservation equation~(\ref{eq-etot}) for the total energy leads to the following balance equation for the internal energy density $e=\epsilon-\rho{\bf v}^2/2$,
\be
\partial_t\, e + \pmb{\nabla}\cdot(e {\bf v}+\pmb{\cal J}_q) = -\pmb{\mathsf P}:\pmb{\nabla}{\bf v}  \, ,
\label{eq-e}
\ee
where $\pmb{\mathsf P}=p\, \pmb{\mathsf I} + \pmb{\mathsf \Pi}$.
Since the macrofield $e({\bf r},t)$ does not depend on the Eulerian angles $\pmb{\alpha}$, equation~(\ref{eq-e}) can be expressed in terms of the time derivative~(\ref{d-dt}) into the following form,
\be
\frac{\dd e}{\dd t} = -(e+p) \, \pmb{\nabla}\cdot{\bf v} - \pmb{\nabla}\cdot\pmb{\cal J}_q -\pmb{\mathsf\Pi}:\pmb{\nabla}{\bf v} \, .
\label{eq-e-ddt}
\ee
Similarly, equation~(\ref{eq-n_k}) can be rewritten as
\be
\frac{\dd n_\varkappa}{\dd t} = - n_\varkappa\, \pmb{\nabla}\cdot{\bf v} - \pmb{\nabla}\cdot\pmb{\cal J}_\varkappa + \sum_{r=1}^{\mathscr R} \nu_{\varkappa r} \, w_r \, .
\label{eq-n_k-ddt}
\ee

Furthermore, if the time derivative~(\ref{d-dt}) is applied to the colloidal distribution function $f_{\rm C}({\bf r},\pmb{\alpha},t)$, its local conservation equation~(\ref{eq-C-3}) with its translational and rotational current densities~(\ref{J_Ct-J_Cr-adv-diss}) reads
\be
\frac{\dd f_{\rm C}}{\dd t} = -f_{\rm C}\, \pmb{\nabla}\cdot{\bf v} - \pmb{\nabla}\cdot\pmb{\mathscr J}_{\rm Ct} - \frac{1}{\sqrt{g}} \, \frac{\partial}{\partial\pmb{\alpha}}\cdot\left(\sqrt{g}\, \pmb{\mathscr J}_{\rm Cr}\right) ,
\label{eq-C-ddt}
\ee
because
\be
\frac{\partial}{\partial\pmb{\alpha}}\cdot\left(\sqrt{g}\, \pmb{\upsilon}_{\rm r}\right) = 0 \, .
\label{dv_r-dalpha}
\ee
Indeed, using the definition~(\ref{v_r}) and the matrix~(\ref{matrix-N}), we have
\be
\pmb{\upsilon}_{\rm r} = \pmb{\mathsf N}^{{\rm T}-1}\cdot\frac{\pmb{\omega}}{2}
= \frac{1}{2}
\left(
\begin{array}{ccc}
-\sin\phi & \cos\phi & 0 \\
-\cot\theta \, \cos\phi & -\cot\theta \, \sin\phi & 1 \\
\frac{\cos\phi}{\sin\theta} & \frac{\sin\phi}{\sin\theta} & 0
\end{array}
\right)\cdot
\left(
\begin{array}{c}
\omega^x \\
\omega^y \\
\omega^z
\end{array}
\right) ,
\ee
so that
\be
\frac{\partial}{\partial\pmb{\alpha}}\cdot\left(\sqrt{g}\, \pmb{\upsilon}_{\rm r}\right) = \frac{1}{2} \, \left(\partial_\theta \ \partial_\phi \ \partial_\psi\right) \cdot
\left(
\begin{array}{c}
-\omega^x \sin\theta \, \sin\phi +\omega^y \sin\theta \, \cos\phi \\
-\omega^x \cos\theta \, \cos\phi -\omega^y \cos\theta \, \sin\phi  + \omega^z \, \sin\theta \\
\omega^x \, \cos\phi + \omega^y \sin\phi 
\end{array}
\right) = 0 \, ,
\ee
because the fluid vorticity $\pmb{\omega}=\pmb{\nabla}\times{\bf v}$ does not depend on the Eulerian angles $\pmb{\alpha}$, which proves equation~(\ref{dv_r-dalpha}).

Next, we replace equations~(\ref{eq-e-ddt}), (\ref{eq-n_k-ddt}), and~(\ref{eq-C-ddt}) into equation~(\ref{Gibbs-ddt}), and using the local Euler thermodynamic relation~(\ref{Euler}) and the identities
\be
X \, \pmb{\nabla}\cdot\pmb{\cal J}_x = \pmb{\nabla}\cdot\left( X \, \pmb{\cal J}_x\right) - \pmb{\cal J}_x\cdot\pmb{\nabla}X
\qquad\mbox{and}\qquad
\int \zeta_{\rm C} \, \frac{1}{\sqrt{g}} \, \frac{\partial}{\partial\pmb{\alpha}}\cdot\left(\sqrt{g}\, \pmb{\mathscr J}_{\rm Cr}\right) \, \ddo = - \int \pmb{\mathscr J}_{\rm Cr}\cdot \frac{\partial\zeta_{\rm C}}{\partial\pmb{\alpha}} \, \ddo \, ,
\ee
we find the following equation for the entropy density,
\be
\frac{\dd s}{\dd t} = -s \, \pmb{\nabla}\cdot{\bf v} - \pmb{\nabla}\cdot\pmb{\cal J}_s + \sigma_s
\label{eq-s-ddt}
\ee
with the entropy dissipative current density
\be
\pmb{\cal J}_s = \frac{1}{T} \, \pmb{\cal J}_q - \sum_{\varkappa=0}^{\mathscr M} \frac{\mu_\varkappa}{T} \,  \pmb{\cal J}_\varkappa - \int \frac{\zeta_{\rm C}}{T} \, \pmb{\mathscr J}_{\rm Ct} \, \ddo
\label{J_s-0}
\ee
and the entropy production rate density
\be
\sigma_s = \pmb{\cal J}_q \cdot \pmb{\nabla}\frac{1}{T} - \frac{1}{T} \, \pmb{\mathsf\Pi}:\pmb{\nabla}{\bf v} - \sum_{\varkappa=0}^{\mathscr M} \pmb{\cal J}_\varkappa \cdot \pmb{\nabla}\frac{\mu_\varkappa}{T} - \frac{1}{T} \sum_{\varkappa=0}^{\mathscr M} \sum_{r=1}^{\mathscr R} \mu_\varkappa \, \nu_{\varkappa r} \, w_r -  \int \pmb{\mathscr J}_{\rm Ct} \cdot \pmb{\nabla}\frac{\zeta_{\rm C}}{T} \, \ddo -  \int \pmb{\mathscr J}_{\rm Ct} \cdot \frac{\partial}{\partial\pmb{\alpha}} \frac{\zeta_{\rm C}}{T} \, \ddo \, .
\label{sigma-s-0}
\ee

We note that equation~(\ref{eq-s-ddt}) is equivalent to the local balance equation~(\ref{eq-s}) for the entropy density, because of equation~(\ref{d-dt}).  Moreover, the dissipative part of the pressure tensor and the tensor of velocity gradients can split into their traceless part and the unit tensor multiplied by one third of their trace, leading to the following decomposition
\be
-\frac{1}{k_{\rm B}T} \, \pmb{\mathsf\Pi}:\pmb{\nabla}{\bf v} = 
{\stackrel{\circ}{\pmb{\mathsf\Pi}}} : \; \stackrel{\circ}{{\pmb{\mathsf A}}}_{\bf g} + \, \Pi \, {\cal A}_{\bf g}
\label{traceless-trace-decomp}
\ee
in terms of the affinities associated with linear momentum transport and defined in the two last lines of table~\ref{Tab1}.
Finally, using the expression~(\ref{J_0}) for the dissipative current density of the solvent and the other constraint~(\ref{constraint-1}) due to mass conservation, we obtain equation~(\ref{J_s}) for the dissipative current density of entropy and equation~(\ref{sigma_s}) for the entropy production rate density in terms of the relative chemical potentials (\ref{rel-mu_k}) and~(\ref{rel-zeta_C}) and the affinities and current densities of table~\ref{Tab1}, as announced.

\section{Equilibrium thermodynamics of dilute suspensions}
\label{AppD}

The Gibbs free energy of dilute solutions can be calculated using equilibrium statistical mechanics, as shown in Ref.~\cite{LL80}.  In a dilute suspension, the colloidal particles constitute an additional species, having extra degrees of freedom for their orientation.  Accordingly, the Gibbs free energy of the dilute suspension should include a corresponding contribution depending on the distribution function of the colloidal particles.  If $\varkappa=0$ denotes the solvent molecular species and $\varkappa=1,2,\dots,{\mathscr M}$ the solute species, the Gibbs free energy of the dilute suspension is thus given by
\be
G=\int \left[ \mu_0^0 \, n_0 + \sum_{\varkappa=1}^{\mathscr M} n_\varkappa \left( \mu_\varkappa^0 + k_{\rm B}T \, \ln\frac{n_\varkappa}{{\rm e}\, n_0}\right) + \int f_{\rm C} \left( \zeta_{\rm C}^0 + k_{\rm B}T \, \ln\frac{8\pi^2 f_{\rm C}}{{\rm e}\, n_0}\right)  \ddo \right] \ddv \, ,
\label{G-dilute}
\ee
where $n_\varkappa$ are the molecular densities; $\mu_0^0$, $\mu_\varkappa^0$, and $\zeta_{\rm C}^0$ are functions of the solvent density $n_0$, the temperature $T$, and the pressure $p$; and ${\rm e}=2.71828\dots$ is Euler's number.  The expression~(\ref{G-dilute}) holds up to corrections going as the second powers of the solute molecular densities and the colloidal distribution function, which are due to the binary interactions between all the solute species including the colloidal species \cite{LL80}.

The Gibbs free energy can be expressed as $G= \int_{\cal V} g \, \ddv$ in terms of its density
\be
g \equiv e+p - Ts = \sum_{\varkappa=0}^{\mathscr M} \mu_\varkappa \, n_\varkappa + \int \zeta_{\rm C} \, f_{\rm C} \, \ddo \, .
\label{g-density}
\ee
The corresponding local Gibbs differential relation reads
\be
\delta g= \delta p - s \, \delta T + \sum_{\varkappa=0}^{\mathscr M} \mu_\varkappa \, \delta n_\varkappa + \int \zeta_{\rm C} \, \delta f_{\rm C} \, \ddo \, .
\label{Gibbs-g-density}
\ee
Therefore, we can deduce the molecular and colloidal chemical potentials for the dilute suspension as
\bea
&& \mu_0 = \mu_0^0 - \frac{k_{\rm B}T}{n_0} \left( \sum_{\varkappa=1}^{\mathscr M} n_\varkappa + n_{\rm C} \right) , \label{mu_0-dilute}\\
&& \mu_\varkappa = \mu_\varkappa^0 + k_{\rm B}T \, \ln\frac{n_\varkappa}{n_0} 
\qquad\mbox{for}\qquad \varkappa=1,2,\dots,{\mathscr M} \, , \label{mu_k-dilute}\\
&&  \zeta_{\rm C} = \zeta_{\rm C}^0 + k_{\rm B}T \, \ln\frac{8\pi^2 f_{\rm C}}{n_0} \, , \label{zeta_C-dilute}
\eea
up to corrections due to the binary interactions.  The quantity $\mu_0^0$ can be interpreted as the chemical potential of the pure solvent, $\mu_\varkappa^0$ as the chemical potential of the solute species $\varkappa$ in the limit where its density $n_\varkappa$ would be extrapolated to the solvent density $n_0$, and $\zeta_{\rm C}^0$ as the chemical potential of the colloidal species in the limit where its distribution function would be extrapolated to $f_{\rm C}=n_0/(8\pi^2)$.

\vskip 0.2 cm

We note that the Gibbs free energy of a dilute suspension of axisymmetric colloidal particles can be equivalently expressed in terms of the moments of the colloidal distribution function.  Using the expansion (\ref{expansion-f_C-truncated}) and assuming that the reference chemical potential $\zeta_{\rm C}^0$ of the colloidal particles does not depend on their orientation (as expected in the absence of external force field), the Gibbs free energy~(\ref{G-dilute}) becomes
\be
G=\int \left[ \mu_0^0 \, n_0 + \sum_{\varkappa=1}^{\mathscr M} n_\varkappa \left( \mu_\varkappa^0 + k_{\rm B}T \, \ln\frac{n_\varkappa}{{\rm e}\, n_0}\right) + n_{\rm C} \left( \mu_{\rm C}^0 + k_{\rm B}T \, \ln\frac{n_{\rm C}}{{\rm e}\, n_0}\right)  + \frac{3 k_{\rm B}T}{2 n_{\rm C}} \, {\bf p}^2 +\cdots \right] \ddv \, ,
\label{G-dilute-expanded}
\ee
where $\mu_{\rm C}^0 \equiv \int \zeta_{\rm C}^0 \, \ddo/(4\pi)$ and the dots denote higher-order terms.
The expression~(\ref{G-dilute-expanded}) shows that, if there is no orientational order (i.e., if ${\bf p}=0$, $\pmb{\mathsf q}=0$, $\dots$), the colloidal species behave as an additional solute species.

\section{Connection with the work of Brenner and coworkers}
\label{AppE}

In the absence of reactions and molecular concentration gradients, the deterministic force and torque exerted on a colloidal particle of arbitrary shape by a fluid in the laminar regime have been shown by Brenner and coworkers \cite{B63,B64a,B64b,B65,B67,BC72,HB83} to be given by
\be
\left(
\begin{array}{c}
{\bf F} \\
{\bf T}
\end{array}
\right)
=
-\left(
\begin{array}{cc}
\pmb{\mathsf\Gamma}_{\rm t} & \pmb{\mathsf\Gamma}_{\rm c}^{\rm T} \\
\pmb{\mathsf\Gamma}_{\rm c} & \pmb{\mathsf\Gamma}_{\rm r}
\end{array}
\right)
\cdot
\left(
\begin{array}{c}
{\bf V}-{\bf v} \\
\pmb{\Omega} -\frac{\pmb{\omega}}{2}
\end{array}
\right)
-\left(
\begin{array}{c}
\pmb{\mathsf\Psi}_{\rm t}\\
\pmb{\mathsf\Psi}_{\rm r}
\end{array}
\right)
:\pmb{\mathsf S}
+\left(
\begin{array}{c}
{\bf F}_{\rm ext} \\
{\bf T}_{\rm ext}
\end{array}
\right) ,
\label{F-T-Brenner}
\ee
where $\pmb{\mathsf\Gamma}_{\rm t}=\eta_0\pmb{\mathsf K}_{\rm t}$ is a $3\times 3$ symmetric matrix of translational friction, $\pmb{\mathsf\Gamma}_{\rm r}=\eta_0\pmb{\mathsf K}_{\rm r}$ is a $3\times 3$ symmetric matrix of rotational friction, $\pmb{\mathsf\Gamma}_{\rm c}=\eta_0\pmb{\mathsf K}_{\rm c}$ is a $3\times 3$ matrix of coupling between translation and rotation, ${\bf V}$ and $\pmb{\Omega}$ are the particle translational and angular velocities, ${\bf v}$ is the fluid velocity, $\pmb{\omega}=\pmb{\nabla}\times{\bf v}$ is the fluid vorticity, $\pmb{\mathsf\Psi}_{\rm t}=\eta_0\pmb{\mathsf\upphi}$ and $\pmb{\mathsf\Psi}_{\rm r}=\eta_0\pmb{\mathsf\uptau}$ are triadics or rank-three tensors of coupling to the symmetrised velocity gradient tensor $\pmb{\mathsf S}=\frac{1}{2}(\pmb{\nabla}{\bf v}+\pmb{\nabla}{\bf v}^{\rm T})$, and
\be
{\bf F}_{\rm ext} = - \pmb{\nabla} U_{\rm ext}
\qquad\mbox{and}\qquad
{\bf T}_{\rm ext} = - \pmb{\mathsf N}^{-1} \cdot \partial_{\pmb{\alpha}} U_{\rm ext}
\label{F_ext-T_ext}
\ee
are possible external force and torque exerted on the particle, $\eta_0$ being the shear viscosity of the fluid surrounding the particle.

Assuming that the particle is dragged by the fluid (and the possible external force field) and that its movements are overdamped and noiseless, the resultants of all the forces and torques acting on the particle are equal to zero, i.e., ${\bf F}=0$ and ${\bf T}=0$ in equation~(\ref{F-T-Brenner}).  This implies that, after the inversion of the $6\times 6$ matrix of friction, the translational and angular velocities of the particle can be expressed as
\be
\left(
\begin{array}{c}
{\bf V} \\
\pmb{\Omega}
\end{array}
\right)
=
\left(
\begin{array}{c}
{\bf v} \\
\frac{\pmb{\omega}}{2}
\end{array}
\right)
+\left(
\begin{array}{cc}
\pmb{\mathsf M}_{\rm t} & \pmb{\mathsf M}_{\rm c}^{\rm T} \\
\pmb{\mathsf M}_{\rm c} & \pmb{\mathsf M}_{\rm r}
\end{array}
\right)
\cdot
\left(
\begin{array}{c}
{\bf F}_{\rm ext} \\
{\bf T}_{\rm ext}
\end{array}
\right)
-
\left(
\begin{array}{c}
\pmb{\mathsf A}\\
\pmb{\mathsf B}
\end{array}
\right)
:\pmb{\mathsf S}
\label{V-Omega-Brenner}
\ee
in terms of the mobility matrices
\bea
&& \pmb{\mathsf M}_{\rm t} = \left( \pmb{\mathsf\Gamma}_{\rm t} - \pmb{\mathsf\Gamma}_{\rm c}^{\rm T}\cdot\pmb{\mathsf\Gamma}_{\rm r}^{-1}\cdot \pmb{\mathsf\Gamma}_{\rm c}\right)^{-1} \, , \label{M_t}\\
&& \pmb{\mathsf M}_{\rm r} = \left( \pmb{\mathsf\Gamma}_{\rm r} - \pmb{\mathsf\Gamma}_{\rm c}\cdot\pmb{\mathsf\Gamma}_{\rm t}^{-1}\cdot \pmb{\mathsf\Gamma}_{\rm c}^{\rm T}\right)^{-1} \, , \label{M_r}\\
&& \pmb{\mathsf M}_{\rm c} = \left( \pmb{\mathsf\Gamma}_{\rm c}^{\rm T} - \pmb{\mathsf\Gamma}_{\rm t}\cdot\pmb{\mathsf\Gamma}_{\rm c}^{-1}\cdot \pmb{\mathsf\Gamma}_{\rm r}\right)^{-1} \, , \label{M_c}
\eea
and the following rank-three tensors,
\bea
&& \pmb{\mathsf A} = \pmb{\mathsf M}_{\rm t}\cdot \pmb{\mathsf\Psi}_{\rm t} + \pmb{\mathsf M}_{\rm c}^{\rm T}\cdot \pmb{\mathsf\Psi}_{\rm r} \, , \label{3tensor-A}\\
&& \pmb{\mathsf B} = \pmb{\mathsf M}_{\rm c}\cdot \pmb{\mathsf\Psi}_{\rm t} + \pmb{\mathsf M}_{\rm r}\cdot \pmb{\mathsf\Psi}_{\rm r} \, , \label{3tensor-B}
\eea
such that $\pmb{\mathsf A}=({\mathsf A}^{ijk})$ and $\pmb{\mathsf B}=({\mathsf B}^{ijk})$ \cite{BC72}.

\vskip 0.2 cm

In Cartesian coordinates, the six-dimensional dissipative current density of the colloidal particles can thus be expressed as
\be
{\bf J}_{\rm C} = 
\left(
\begin{array}{c}
{\bf V} \\
\pmb{\Omega}
\end{array}
\right)
f_{\rm C}
-
\left(
\begin{array}{cc}
\pmb{\mathsf D}_{\rm t} & \pmb{\mathsf D}_{\rm c}^{\rm T} \\
\pmb{\mathsf D}_{\rm c} & \pmb{\mathsf D}_{\rm r}
\end{array}
\right)
\cdot
\left(
\begin{array}{c}
\pmb{\nabla} f_{\rm C} \\
\partial_{\pmb{\chi}} f_{\rm C}
\end{array}
\right) ,
\qquad\mbox{where}\qquad
\pmb{\mathsf D}_{\rm x} = k_{\rm B}T \, \pmb{\mathsf M}_{\rm x}
\qquad\mbox{for}\quad
{\rm x}={\rm t},{\rm r},{\rm c}
\label{J_C-Brenner}
\ee
are the diffusion matrices corresponding to the mobility matrices~(\ref{M_t})-(\ref{M_c}).
This result is a consequence of the fact that the current density should be equal to zero at equilibrium, when the distribution function is given by Boltzmann's distribution $f_{\rm C}\sim\exp(-\beta U_{\rm ext})$ with the inverse temperature $\beta=(k_{\rm B}T)^{-1}$ in a fluid at rest where ${\bf v}=0$, $\pmb{\omega}=0$, and $\pmb{\mathsf S}=0$ \cite{PM53,B65}.

\vskip 0.2 cm

According to equation~(\ref{J_C-Cartesian-curvilinear}), we thus find that, in the curvilinear coordinates of Eulerian angles, the colloidal current density is given by
\be
\pmb{\mathfrak J}_{\rm C} = \pmb{\mathfrak V} \, f_{\rm C} - \pmb{\mathscr D} \cdot
\left(
\begin{array}{c}
\pmb{\nabla} f_{\rm C} \\
\partial_{\pmb{\alpha}} f_{\rm C}
\end{array}
\right)
\label{J_C-Brenner-curvilin}
\ee
with
\be
\pmb{\mathfrak V} =
\left(
\begin{array}{c}
\pmb{\mathfrak V}_{\rm t} \\
\pmb{\mathfrak V}_{\rm r}
\end{array}
\right)
=
\pmb{\mathsf\Theta}^{\rm T} \cdot
\left(
\begin{array}{c}
{\bf V} \\
\pmb{\Omega}
\end{array}
\right)
\label{mathfrak-V}
\ee
and
\be
\pmb{\mathscr D}
=
\left(
\begin{array}{cc}
\pmb{\mathscr D}_{\rm t} & \pmb{\mathscr D}_{\rm c}^{\rm T} \\
\pmb{\mathscr D}_{\rm c} & \pmb{\mathscr D}_{\rm r}
\end{array}
\right)
=
\pmb{\mathsf\Theta}^{\rm T}\cdot
\left(
\begin{array}{cc}
\pmb{\mathsf D}_{\rm t} & \pmb{\mathsf D}_{\rm c}^{\rm T} \\
\pmb{\mathsf D}_{\rm c} & \pmb{\mathsf D}_{\rm r}
\end{array}
\right)
\cdot
\pmb{\mathsf\Theta}
\label{mathscr-D}
\ee
in terms of the $6\times 6$ matrix $\pmb{\mathsf\Theta}$ introduced in equation~(\ref{J_C-Cartesian-curvilinear}), giving the expressions of the translational and rotational diffusion tensors and the following translational and rotational velocity vectors in the absence of external fields,
\bea
&& \pmb{\mathfrak V}_{\rm t} = {\bf v} + \pmb{\mathscr V}_{\rm t} = {\bf v} - \pmb{\mathsf A}:\pmb{\mathsf S} \, , \label{mathfrak-V-t}\\
&& \pmb{\mathfrak V}_{\rm r} = \pmb{\upsilon}_{\rm r} + \pmb{\mathscr V}_{\rm r} = \pmb{\upsilon}_{\rm r} - \pmb{\mathsf N}^{{\rm T}-1} \cdot \pmb{\mathsf B}:\pmb{\mathsf S} \, , \label{mathfrak-V-r}
\eea
with $\pmb{\upsilon}_{\rm r}$ defined by equation~(\ref{v_r}), provided that the affinities of reactions and molecular transports are equal to zero, i.e., if ${\cal A}_r=0$ and $\pmb{\cal A}_\varkappa=0$ \cite{B67,BC72}.  Therefore, equations~(\ref{mathfrak-V-t}) and~(\ref{mathfrak-V-r}) give the expressions~(\ref{V_t^g}) and~(\ref{V_r^g}) for the contributions of the gradients of the fluid velocity to the translational and rotational dissipative drift velocities of the colloidal particles.

\section{Expansion of the colloidal distribution function}
\label{AppF}

\subsection{Expansion in series of orientational-order tensors}

Here, the colloidal particles are assumed to be axisymmetric, their orientation being determined by the unit vector ${\bf u}=(\sin\theta\cos\phi,\sin\theta\sin\phi,\cos\theta)$, which belongs to the unit sphere ${\mathbb S}^2$.  Accordingly, the distribution function depends only on the two angles $\pmb{\alpha}=(\theta,\phi)$ and the element of integration is $\ddo=\sin\theta\, \dd\theta \, \dd\phi$, such that $\int_{{\mathbb S}^2} \ddo = 4\pi$.  In this case, the distribution function can be expanded  using Buckingham's formula \cite{Bu67,T11,GV23} as
\be
f_{\rm C}({\bf r},\pmb{\alpha},t) = \frac{1}{4\pi} \sum_{\ell=0}^{\infty} \frac{(2\ell + 1) ! ! }{\ell \, !} \, {\mathsf T}_\ell^{i_1\cdots i_\ell} (\pmb{\alpha}) \, {\mathsf t}_\ell^{i_1\cdots i_\ell} ({\bf r},t)
\qquad\mbox{with}\qquad
{\mathsf t}_\ell^{i_1\cdots i_\ell} ({\bf r},t) = \int {\mathsf T}_\ell^{i_1\cdots i_\ell} (\pmb{\alpha})\, f_{\rm C}({\bf r},\pmb{\alpha},t) \, \ddo \, ,
\label{Buckingham_formula}
\ee
where ${\mathsf T}_\ell^{i_1\cdots i_\ell} (\pmb{\alpha})$ are rank-$\ell$ tensors that are totally symmetric under the permutations of the $\ell$ indices, traceless with respect to any pair of indices, and satisfying
\be
\hat{\cal L}_{\rm r} {\mathsf T}_\ell^{i_1\cdots i_\ell} (\pmb{\alpha}) = - \ell(\ell +1) \, {\mathsf T}_\ell^{i_1\cdots i_\ell} (\pmb{\alpha}) \, ,
\label{Laplacian-T}
\ee
i.e., being the eigenvectors of the rotational Laplacian operator~(\ref{rot-Laplacian-2}).  Explicitly, the expansion reads
\be
f_{\rm C} = \frac{1}{4\pi}\left( n_{\rm C} + 3 \, u^i \, p^i + \frac{15}{2} \, {\mathsf Q}_{\bf u}^{ij} \, {\mathsf q}^{ij} + \frac{35}{2} \, {\mathsf R}_{\bf u}^{ijk} \, {\mathsf r}^{ijk} + \frac{315}{8} \, {\mathsf S}_{\bf u}^{ijkl} \, {\mathsf s}^{ijkl} + \cdots \right) ,
\label{f_C-expansion-axisymm}
\ee
where
\bea
&&\ell = 0: \qquad n_{\rm C} = \int f_{\rm C} \, \ddo \, , \label{l=0}\\
&&\ell = 1: \qquad p^i = \int u^i \, f_{\rm C} \, \ddo \, , \label{l=1}\\
&&\ell = 2: \qquad {\mathsf q}^{ij} = \int {\mathsf Q}_{\bf u}^{ij} \, f_{\rm C} \, \ddo
\qquad\mbox{with}\qquad
{\mathsf Q}_{\bf u}^{ij} = u^i u^j - \frac{1}{3}\, \delta^{ij} \, , \label{l=2}\\
&&\ell = 3: \qquad {\mathsf r}^{ijk} = \int {\mathsf R}_{\bf u}^{ijk} \, f_{\rm C} \, \ddo
\qquad\mbox{with}\qquad
{\mathsf R}_{\bf u}^{ijk} = u^i u^j u^k - \frac{1}{5} \left( u^i\delta^{jk} + u^j \delta^{ik} + u^k \delta^{ij}\right) , \label{l=3}\\
&&\ell = 4: \qquad {\mathsf s}^{ijkl} = \int {\mathsf S}_{\bf u}^{ijkl} \, f_{\rm C} \, \ddo 
\qquad\mbox{with}\qquad\nonumber\\
&&\qquad {\mathsf S}_{\bf u}^{ijkl} = u^i u^j u^k u^l - \frac{1}{7} \left( u^i u^j \delta^{kl} + u^i u^k \delta^{jl} + u^i u^l \delta^{jk} + u^j u^k \delta^{il} + u^j u^l \delta^{ik} + u^k u^l \delta^{ij} \right) \nonumber\\
&&\qquad\qquad\qquad\qquad\quad\  + \frac{1}{35} \left(\delta^{ij} \delta^{kl} + \delta^{ik} \delta^{jl} + \delta^{il} \delta^{jk} \right)  , \label{l=4}\\
&&\qquad\qquad\qquad\vdots \nonumber
\eea

We note that the expansion~(\ref{Buckingham_formula}) or~(\ref{f_C-expansion-axisymm}) corresponds to an expansion in a series of spherical harmonics, which can be generalised to the distribution function of non-axisymmetric particles as in equation~(\ref{Wigner-series}) \cite{BP76,T11}.

\subsection{Application to angular integrals}

We have the following identity,
\be
\int g_{ij} \; {\rm grad}_{\rm r}^i {\bf u} \; {\rm grad}_{\rm r}^j f_{\rm C} \, \ddo = \int g^{ij} \, \frac{\partial{\bf u}}{\partial\alpha^i} \, \frac{\partial f_{\rm C}}{\partial\alpha^j} \, \ddo = - \int (\hat{\cal L}_{\rm r}{\bf u})\, f_{\rm C}\, \ddo = 2\, {\bf p} \, ,
\label{int-grad_u-grad_f}
\ee
which is obtained using the expression~(\ref{2_components-grad_r}) for the rotational gradient, the inverse of the metric~(\ref{2-matrix-g}), an integration by parts, the fact that the unit vector $\bf u$ is an eigenvector with $\ell=1$ of the rotational Laplacian operator~(\ref{rot-Laplacian-2}), i.e., $\hat{\cal L}_{\rm r}{\bf u}=-2{\bf u}$, and the definition~(\ref{l=1}).

We also have
\be
\int \frac{1}{f_{\rm C}} \left(\nabla^i f_{\rm C}\right)^2 \, \ddo = \frac{1}{n_{\rm C}} \left[ \left(\nabla^i n_{\rm C}\right)^2 + 3 \left(\nabla^i p^j - p^j\frac{\nabla^i n_{\rm C}}{n_{\rm C}}\right)^2 \right] + \cdots
\label{int-nabla_f-nabla_f}
\ee
and
\be
\int \frac{g^{ij}}{f_{\rm C}} \, \frac{\partial f_{\rm C}}{\partial\alpha^i} \, \frac{\partial f_{\rm C}}{\partial\alpha^j} \, \ddo = \frac{6}{n_{\rm C}} \, {\bf p}^2 + \cdots \, ,
\label{int-grad_f-grad_f}
\ee
where the dots denote higher-order terms.

\subsection{Application to equilibrium thermodynamics}

The expansion~(\ref{f_C-expansion-axisymm}) can also be used to express the local Gibbs and Euler thermodynamic relations (\ref{Gibbs}) and~(\ref{Euler}) as follows,
\be
\delta s = \frac{1}{T} \, \delta e - \sum_{\varkappa = 0}^{\mathscr M} \frac{\mu_\varkappa}{T} \, \delta n_\varkappa - \frac{\mu_{\rm C}}{T} \, \delta n_{\rm C} - \frac{\varphi_{\rm C}^i}{T} \, \delta p^i - \frac{\vartheta_{\rm C}^{ij}}{T} \, \delta{\mathsf q}^{ij} - \cdots
\label{Gibbs-moments}
\ee
and
\be
s = \frac{1}{T} \left( e + p \right)  - \sum_{\varkappa = 0}^{\mathscr M} \frac{\mu_\varkappa}{T} \, n_\varkappa - \frac{\mu_{\rm C}}{T} \, n_{\rm C} - \frac{\varphi_{\rm C}^i}{T} \, p^i - \frac{\vartheta_{\rm C}^{ij}}{T} \, {\mathsf q}^{ij} - \cdots
\label{Euler-moments}
\ee
in terms of the moments~(\ref{l=0})-(\ref{l=4}) of the distribution function and conjugate macrofields defined as
\be
\mu_{\rm C} \equiv \frac{1}{4\pi} \int \zeta_{\rm C} \, \ddo \, , \qquad
\varphi_{\rm C}^i \equiv \frac{3}{4\pi} \int u^i \, \zeta_{\rm C} \, \ddo \, , \qquad
\vartheta_{\rm C}^{ij} \equiv \frac{15}{8\pi} \int {\mathsf Q}_{\bf u}^{ij} \, \zeta_{\rm C} \, \ddo \, , \qquad \dots
\label{conjugate-fields}
\ee
Similar conjugate macrofields have been considered in Ref.~\cite{JGS18}.
In dilute suspensions where the Gibbs free energy~(\ref{G-dilute-expanded}) holds, the first of them is given by $(\varphi_{\rm C}^i )=\pmb{\varphi}_{\rm C}=3(k_{\rm B}T/n_{\rm C}){\bf p} + \cdots$, up to higher-order terms.

\section{Evolution equations for the macrofields of polar and nematic orders}
\label{AppG}

For an incompressible suspension where $\pmb{\nabla}\cdot{\bf v}=0$, the local conservation equation~(\ref{eq-C-3}) for the colloidal particles with the current densities~(\ref{J_Ct-J_Cr-adv-diss}) reads
\be
\partial_t \, f_{\rm C} +{\bf v}\cdot\pmb{\nabla} f_{\rm C} + {\rm div}_{\rm r}\left(f_{\rm C} \, \pmb{\upsilon}_{\rm r}\right) = - \pmb{\nabla}\cdot\pmb{\mathscr J}_{\rm Ct} - {\rm div}_{\rm r}\pmb{\mathscr J}_{\rm Cr} \, ,
\label{eq-C-4}
\ee
as expressed in terms of the rotational fluid velocity~(\ref{v_r}).  In order to derive the evolution equation of one of the moments of the distribution function, this equation should be multiplied by the corresponding orientational tensor and integrated over the Eulerian angles.  Here below, this systematic procedure is carried out for the first two moments of polar and nematic orders.  The left-hand side of equation~(\ref{eq-C-4}) gives the contribution from reversible translational and rotational advections and its right-hand side the contribution from the dissipative part of the current densities.  In the following, we use the identity
\be
\epsilon^{ijk} \, \omega^k = \nabla^i v^j - \nabla^j v^i \, ,
\qquad\mbox{where}\qquad
\omega^k = \epsilon^{klm} \nabla^l v^m 
\label{omega-antisym}
\ee
is the fluid vorticity.

\subsection{Evolution equation for the vector of polar order}

Polar order is characterised by the vector~(\ref{l=1}).  Taking its time derivative and using equation~(\ref{eq-C-4}), we obtain the following equation.  Its left-hand side due to reversible advection is given by
\be
\frac{D p^i}{Dt} \equiv \partial_t \, p^i + {\bf v}\cdot\pmb{\nabla} p^i + \frac{1}{2} \left(\nabla^i v^j-\nabla^j v^i\right) p^j 
\label{DpDt}
\ee
and its right-hand side by
\bea
\frac{D p^i}{Dt} &=& -\nabla^j \Bigg\{ V_{\rm sd} \left({\mathsf q}^{ij} + \frac{1}{3} \, n_{\rm C} \, \delta^{ij}\right) + p^i \nabla^j\left(\sum_\varkappa \xi_\varkappa n_\varkappa\right) \nonumber\\
&& + \left[ {\mathsf r}^{ijk} + \frac{1}{5}\left( p^j \delta^{ik} + p^k \delta^{ij}-\frac{2}{3}\, p^i \delta^{jk}\right)\right] \nabla^k \left(\sum_\varkappa \varepsilon_\varkappa n_\varkappa\right)\Bigg\}
+ D_{\rm t} \nabla^2 p^i \nonumber\\
&& - \left({\mathsf q}^{ij} - \frac{2}{3} \, n_{\rm C} \, \delta^{ij}\right) \nabla^j\left(\sum_\varkappa \lambda_\varkappa n_\varkappa\right) - 2 D_{\rm r} \, p^i \, .
\label{eq-p}
\eea
The left-hand side~(\ref{DpDt}) has the known form \cite{MJRLPRA13,JGS18}.
The last term of the right-hand side~(\ref{eq-p}) expresses the damping of the polar order parameter due to the rotational diffusion of the colloidal particles.

\subsection{Evolution equation for the tensor of nematic order}

This order is characterised by the tensor~(\ref{l=2}) and its time evolution can be directly deduced from equation~(\ref{eq-C-4}).  The dynamical equation ruling this tensor is formed by the following left-hand side due to reversible advection,
\be
\frac{D {\mathsf q}^{ij}}{Dt} \equiv \partial_t \, {\mathsf q}^{ij} + {\bf v}\cdot\pmb{\nabla} {\mathsf q}^{ij} + \frac{1}{2} \left(\nabla^i v^k-\nabla^k v^i\right){\mathsf q}^{kj} + \frac{1}{2} \left(\nabla^j v^k-\nabla^k v^j\right){\mathsf q}^{ki}
\label{DqDt}
\ee
and its right-hand side is given by
\bea
\frac{D {\mathsf q}^{ij}}{Dt} &=& -\nabla^k \Bigg\{ V_{\rm sd} \left[{\mathsf r}^{ijk} + \frac{1}{5} \left( p^i \delta^{jk} + p^j \delta^{ik}-\frac{2}{3}\, p^k \delta^{ij}\right)\right] + {\mathsf q}^{ij}\, \nabla^k\left(\sum_\varkappa \xi_\varkappa n_\varkappa\right) \nonumber\\
&& + \bigg[ {\mathsf s}^{ijkl} + \frac{1}{7}\left( {\mathsf q}^{ik} \delta^{jl} + {\mathsf q}^{il} \delta^{jk} + {\mathsf q}^{jk} \delta^{il} + {\mathsf q}^{jl} \delta^{ik} - \frac{4}{3} \, {\mathsf q}^{ij} \delta^{kl} - \frac{4}{3} \, {\mathsf q}^{kl} \delta^{ij} \right)
\nonumber\\
&& + \frac{n_{\rm C}}{15} \left(\delta^{ik}\delta^{jl}+\delta^{il}\delta^{jl} - \frac{2}{3}\, \delta^{ij}\delta^{kl}\right)\bigg] \nabla^l \left(\sum_\varkappa \varepsilon_\varkappa n_\varkappa\right)\Bigg\}
+ D_{\rm t} \nabla^2 {\mathsf q}^{ij} \nonumber\\
&& - \left[ 2 \, {\mathsf r}^{ijk} + \frac{3}{5} \left( p^i \delta^{jk} + p^j \delta^{ik}-\frac{2}{3}\, p^k \delta^{ij}\right) \right]
\nabla^k\left(\sum_\varkappa \lambda_\varkappa n_\varkappa\right) - 6 D_{\rm r} \, {\mathsf q}^{ij} \, .
\label{eq-q}
\eea
The left-hand side~(\ref{DqDt}) is also known \cite{JGS18}.
The nematic order parameter is damped by the rotational diffusion of the colloidal particles according to the last term of the right-hand side~(\ref{eq-q}).

\section{Inequalities}
\label{AppH}

\subsection{Bound on the self-diffusiophoretic parameter}

In the case where the gradients of velocity and molecular densities are equal to zero, the entropy production rate density~(\ref{sigma_s-sph}) can be written as
\be
\frac{1}{k_{\rm B}} \, \sigma_s = \int f_{\rm C} 
\left(
\begin{array}{cc}
{\cal A}_{\rm r} & -\frac{\pmb{\nabla} f_{\rm C}}{f_{\rm C}} \end{array}
\right)
\cdot
\left(
\begin{array}{cc}
D_{\rm rxn} & \chi \, D_{\rm rxn} \, {\bf u} \\
\chi \, D_{\rm rxn} \, {\bf u} & D_{\rm t} \, \pmb{\mathsf I}
\end{array}
\right)
\cdot
\left(
\begin{array}{c}
 {\cal A}_{\rm r} \\
 -\frac{\pmb{\nabla} f_{\rm C}}{f_{\rm C}} \end{array}
\right)
\ddo
+ D_{\rm r} \int \frac{g^{ij} }{f_{\rm C}}\, \frac{\partial f_{\rm C}}{\partial\alpha^i}\, \frac{\partial f_{\rm C}}{\partial\alpha^j}\, \ddo \ge 0 \, ,
\label{sigma_s-sph-Ar-C}
\ee
where $D_{\rm rxn}$ is the chemical reaction diffusivity introduced in equation~(\ref{Aff-AB-eq}), $\chi$ is the self-diffusiophoretic parameter~(\ref{chi}), and $D_{\rm t}$ and $D_{\rm r}$ are the translational and rotational diffusion coefficients of the colloidal particles, respectively.  The first term of equation~(\ref{sigma_s-sph-Ar-C}) gives the entropy production rate density of self-diffusiophoresis propelling the particles by the reaction of affinity ${\cal A}_{\rm r}$ and the second term is due to rotational Brownian motion.

This entropy production rate density is non-negative under the conditions that
\be
D_{\rm t} \ge \chi^2 \, D_{\rm rxn}
\qquad\mbox{and}\qquad
D_{\rm r} \ge 0 \, .
\label{inequalities-activ}
\ee

We note that we here recover the same $4\times 4$ matrix of linear response as in Ref.~\cite{GK17}, where the role of the affinity $\pmb{\mathscr A}_{\rm Cr}=-\pmb{\nabla} f_{\rm C}/f_{\rm C}$ was played by the mechanical affinity due to an external force.

\subsection{Bound on the diffusiophoretic parameters}

In the other case where there is no reaction (i.e., ${\cal A}_{\rm r}=0$), the colloidal distribution is isotropic [i.e., $f_{\rm C}=n_{\rm C}/(4\pi)$], and no velocity gradients, the entropy production rate density~(\ref{sigma_s-sph}) reduces to
\bea
\frac{1}{k_{\rm B}} \, \sigma_s &=& \sum_{\varkappa=1}^{\mathscr M} D_\varkappa \, \frac{(\nabla^i n_\varkappa)^2}{n_\varkappa} - \, 2 \,  \sum_{\varkappa=1}^{\mathscr M} \xi_\varkappa \, \nabla^i n_\varkappa \, \nabla^i n_{\rm C}  + D_{\rm t} \, \frac{(\nabla^i n_{\rm C})^2}{n_{\rm C}} \nonumber\\
&=& \sum_{\varkappa=1}^{\mathscr M} \frac{D_\varkappa}{n_\varkappa} \, \left(\nabla^i n_\varkappa- \xi_\varkappa \, \frac{n_\varkappa}{D_\varkappa} \, \nabla^i n_{\rm C}\right)^2
+\left( \frac{D_{\rm t}}{n_{\rm C}}- \sum_{\varkappa=1}^{\mathscr M} \xi_\varkappa^2 \, \frac{n_\varkappa}{D_\varkappa}\right) (\nabla^i n_{\rm C})^2 \ge 0  \, ,
\label{sigma_s-sph-n_k-n_C}
\eea
where $D_\varkappa$ is the diffusion coefficient of molecular species $\varkappa$ and $\xi_\varkappa$ the diffusiophoretic parameter given in equation~(\ref{xi-eps-lambda}).  The expression~(\ref{sigma_s-sph-n_k-n_C}) is non-negative under the conditions that
\be
\frac{D_{\rm t}}{n_{\rm C}}  \ge \sum_{\varkappa=1}^{\mathscr M} \xi_\varkappa^2 \, \frac{n_\varkappa}{D_\varkappa}
\qquad\mbox{and}\qquad
D_\varkappa > 0 \, .
\label{inequalities-activ}
\ee

Such reasonings may also be considered to obtain similar bounds for the parameters $\varepsilon_\varkappa$, $\lambda_\varkappa$, and $\varpi\, k_\varkappa$.

\vskip 0.2 cm

All these inequalities justify the conditions~(\ref{inequalities}).


\end{document}